\documentclass[12pt,a4paper]{article}
\pdfoutput=1 % for JHEP
\usepackage{amsmath,amssymb,cite,slashed,tabularx}
\usepackage{subfigure}
\usepackage{graphicx,color}

\allowdisplaybreaks

\numberwithin{equation}{section} % Eq.(Sec.eq.)
\def\beq#1\eeq{\begin{align}#1\end{align}}

\usepackage{caption}
\definecolor{BlueViolet}{rgb}{0.2, 0.00, 0.7}
\definecolor{Blue}{rgb}{0.15, 0.00, 0.9}
\usepackage[colorlinks=true,linkcolor=Blue,citecolor=Blue,urlcolor=BlueViolet]{hyperref}

\begin{document}

\begin{titlepage}

\begin{center}

\hfill KEK--TH--2170\\

\vskip .35in

{\Large\bf
 Nuclear EDM from SMEFT flavor-changing operator
}

\vskip .4in

{\large
  Motoi Endo and
  Daiki Ueda
}

\vskip 0.25in

{\it Theory Center, IPNS, KEK, Tsukuba, Ibaraki 305-0801, Japan}

\vskip 0.1in

{\it The Graduate University of Advanced Studies (Sokendai),\\
Tsukuba, Ibaraki 305-0801, Japan}

\end{center}

\vskip .3in

\begin{abstract}

We study nuclear electric dipole moments induced by $\Delta F=1$ effective operators in the Standard Model Effective Field Theory.
Such contributions arise through renormalization group evolutions and matching conditions at the electroweak symmetry breaking scale.  
We provide one-loop formulae for the matching conditions. 
We also discuss correlations of these effects with $\Delta F=2$ observables such as $\epsilon_K$ and $\Delta M_{B_d}$.

\end{abstract}
\end{titlepage}

\setcounter{page}{1}
\renewcommand{\thefootnote}{\#\arabic{footnote}}
\setcounter{footnote}{0}

%%%%%%%%%%%%%%%%%%%%%%%%%%%%%%%%%%%%%%%%%%%%%%%%%%%%%%%%
\section{Introduction}
\label{sec:introduction}
%%%%%%%%%%%%%%%%%%%%%%%%%%%%%%%%%%%%%%%%%%%%%%%%%%%%%%%%
Since we have not discovered new particles at the LHC experiment~\cite{LHC}, physics beyond the standard model (SM) is very likely to exist in high energy scale, particularly above the electroweak symmetry breaking (EWSB) scale.
Such a high scale can be probed indirectly through flavor and $CP$ violations. 
In particular, electric dipole moments (EDMs), which are $CP$-violating observables, are one of the most sensitive observables.
Currently, an experimental bound on the EDM of the $^{199}$Hg atom is provided as~\cite{Griffith:2009zz,Graner:2016ses}
\begin{align}
|d_{\rm Hg}| < 6.3 \times 10^{-30}~e\, {\rm cm},
\end{align}
at the 90\% confidence level.
Although theoretical calculations suffer from uncertainties in estimating the Schiff moment, it can constrain NP very severely.
For the nucleon, the experimental bound of the neutron is~\cite{Afach:2015sja}
\begin{align}
|d_n| < 3.0 \times 10^{-26}~e\, {\rm cm},
\end{align}
at the 90\% confidence level.
On the other hand, the indirect limit on the proton EDM is derived from $^{199}$Hg as~\cite{Sahoo:2016zvr}
\begin{align}
|d_p| < 2.1 \times 10^{-25}~e\, {\rm cm}.
\end{align}
In future, several experiments aim to improve the sensitivity by two orders of magnitudes for the neutron EDM~\cite{Strategy:2019vxc,Wurm:2019yfj}. Also, a storage ring experiment is projected to measure the proton EDM at the level of $10^{-29}~e\,{\rm cm}$~\cite{Anastassopoulos:2015ura}.

Although the EDMs are flavor-conserving observables, flavor-violating interactions can contribute to them. 
In the SM, the $W$-boson interactions change quark flavors.
Thus, a class of NP can induce EDMs through quark flavor-changing interactions by exchanging the $W$ boson. 
Such contributions are represented by the Standard Model Effective Field Theory (SMEFT)~\cite{Buchmuller:1985jz,Grzadkowski:2010es,Dedes:2017zog}.
Here, all the SM particles including the electroweak bosons $(W, Z, H)$ and the top quark $(t)$ are retained. 
Above the EWSB, NP contributions to flavor and $CP$ violations are encoded in higher dimensional operators in the SMEFT.
At the EWSB scale, they are matched onto the effective operators in the low-scale effective field theory (LEFT) by integrating out $W$, $Z$, $H$ and $t$.
Low-scale observables such as the EDMs are evaluated by using the LEFT.

In this letter, we study the nuclear EDMs from SMEFT flavor-changing operators.
They are induced by $\Delta F=1$ operators through radiative corrections of the $W$ boson. 
In particular, we will focus on top-quark loop contributions, because they tend to be large due to the large top quark mass (cf.~Ref.~\cite{Endo:2018gdn}). 
The radiative corrections are taken into account by solving the renormalization group equations (RGEs) in the SMEFT~\cite{Jenkins:2013zja,Jenkins:2013wua,Alonso:2013hga}.
In addition, the SMEFT operators are matched onto those in the LEFT at the EWSB scale.
The one-loop matching conditions are necessary, because the contributions of the $\Delta F=1$ operators to EDMs are induced by radiative corrections. 
The one-loop formulae will be provided in this letter. 
Theses operators also contribute to $\Delta F=2$ observables such as $\epsilon_K$ and $\Delta M_d$ through the $W$-boson loops.
Since these observables are sensitive to NP contributions, we will discuss correlation between the contributions to the EDMs and the $\Delta F=2$ observables.

%%%%%%%%%%%%%%%%%%%%%%%%%%%%%%%%%%%%%%%%%%%%%%%%%%%%%%%%
\section{Formula}
\label{sec:formula}
%%%%%%%%%%%%%%%%%%%%%%%%%%%%%%%%%%%%%%%%%%%%%%%%%%%%%%%%

In this section, we provide formulae for evaluating the EDMs induced by flavor-changing operators. 
By decoupling NP particles, their contributions are encoded in higher dimensional operators in the SMEFT.
Then, these operators are evolved by following the RGEs.
Anomalous dimensions in the SMEFT are provided at the one-loop level in Refs.~\cite{Jenkins:2013zja,Jenkins:2013wua,Alonso:2013hga}, and those relevant for the $CP$ and flavor violations are summarized in Ref.~\cite{Endo:2018gdn}.
At the EWSB scale, they are matched to the LEFT operators. 
We provide the one-loop matching formulae between the SMEFT $\Delta F=1$ operators and the LEFT $\Delta F=0$ $CP$-violating operators.

In the SMEFT, the $\Delta F = 1$ effects are encoded into higher dimensional operators, which are defined as~\cite{Grzadkowski:2010es}
\begin{align}
 \mathcal{L}_{\rm eff} = \mathcal{L}_{\textrm{SM}} + \sum_a C_a \mathcal{O}_a,
 \label{eq:SMEFT}
\end{align}
where the first term in the right-hand side is the SM Lagrangian, and the second term represents the higher dimensional operators.
Here, the Lagrangian is invariant under the SM gauge symmetry, and all the SM particles including $W, Z, H$ and $t$ are dynamical. 
In this letter, we consider the operators of the down-type quarks,\footnote{It is straightforward to extend the analysis to $\Delta F=1$ operators of the up-type quarks. In this case, radiative corrections are likely to be dominated by bottom-quark loops.}which correlate with $\Delta F=2$ observables, such as $\epsilon_K$ and $\Delta M_{B_b}$. 
The dimension-six operators which potentially relevant to the EDMs and the $\Delta F=2$ observables are shown as\footnote{
 See Ref.~\cite{Alioli:2017ces} for an extensive study of the SMEFT operator, $(\tilde H^\dagger i\overleftrightarrow{D}_\mu H)(\overline{u}^{i} \gamma^{\mu} d^{j})$, where EDMs and flavor observables are examined. 
 Also, the nucleon/nuclear EDM has been discussed within the context of the SMEFT in Ref.~\cite{Dekens:2013zca,Engel:2013lsa,Aebischer:2018csl}, where flavor-conserving operators have been studied.
}
\begin{align}
 &(\mathcal{O}_{qq}^{(1)})_{ijkl} = 
 (\overline{q}^i \gamma_{\mu} q^j)(\overline{q}^k \gamma_{\mu} q^l),
 \\
 &(\mathcal{O}^{(3)}_{qq})_{ijkl} = 
 (\overline{q}^i\gamma_{\mu}\tau^I q^j)(\overline{q}^k \gamma^{\mu}\tau^I q^l),
 \\
 &(\mathcal{O}^{(1)}_{qd})_{ijkl} = 
 (\overline{q}^i\gamma_{\mu} q^j)(\overline{d}^k \gamma^{\mu} d^l),
 \\
 &(\mathcal{O}^{(8)}_{qd})_{ijkl} = 
 (\overline{q}^i \gamma_{\mu}T^A q^j)(\overline{d}^k \gamma^{\mu}T^A d^l),
 \\
& (\mathcal{O}_{dd})_{ijkl} = 
 (\overline{d}^i\gamma_{\mu} d^j)(\overline{d}^k \gamma^{\mu} d^l),
 \\
& (\mathcal{O}^{(1)}_{Hq})_{ij} = 
 (H^\dagger i\overleftrightarrow{D_\mu} H)(\overline{q}^{i} \gamma^{\mu} q^{j}),
 \\
 &(\mathcal{O}^{(3)}_{Hq})_{ij} = 
 (H^\dagger i\overleftrightarrow{D^I_{\mu}} H)(\overline{q}^{i} \gamma^{\mu}\tau^I q^{j}), 
 \\
& (\mathcal{O}_{Hd})_{ij} = 
 (H^\dagger i\overleftrightarrow{D_{\mu}} H)(\overline{d}^{i} \gamma^{\mu} d^{j}),
 \\
 & (\mathcal{O}^{(1)}_{qu})_{ijkl}=(\bar{q}^i \gamma_{\mu} q^j) (\bar{u}^k \gamma^{\mu}u^l),
 \\
 & (\mathcal{O}^{(8)}_{qu})_{ijkl}= (\bar{q}^i \gamma_{\mu}T^A q^j) (\bar{u}^k \gamma^{\mu} T^A u^l),
 \\
 & (\mathcal{O}^{(1)}_{ud})_{ijkl} = 
 (\overline{u}^i \gamma_{\mu} u^j)(\overline{d}^k \gamma^{\mu} d^l),
 \\
& (\mathcal{O}^{(8)}_{ud})_{ijkl} = 
 (\overline{u}^i \gamma_{\mu} T^A u^j)(\overline{d}^k \gamma^{\mu}T^A d^l),
\end{align}
with the derivative,
\begin{align}
 H^\dagger\overleftrightarrow{D^I_\mu} H = 
 H^{\dagger} \tau^I D_{\mu} H - \left( D_{\mu} H\right)^{\dagger}  \tau^I H,
\end{align}
where $q$ is the ${\rm SU(2)}_L$ quark doublet, $d$ the right-handed down-type quark, $u$ the right-handed up-type quark, and $T^A$ the ${\rm SU(3)}_C$ generator with quark-flavor indices $i,j,k,l$ and an ${\rm SU(2)}_L$ [${\rm SU(3)}_C$] index $I$ [$A$]. 

At the EWSB scale, the SMEFT operators are matched to the LEFT operators. 
The latter operators for EDMs are defined as
\begin{align}
\mathcal{L}_{\rm CPV}=
\sum_{a=1,2,4,5}\sum_i C_a^i \mathcal{O}_a^i
+C_3 \mathcal{O}_3
+\sum_{a=1,2}\sum_{i\neq j}\tilde{C}^{ij}_a \tilde{\mathcal{O}}^{ij}_a
+\frac{1}{2}\sum_{a=3,4}\sum_{i\neq j}\tilde{C}^{ij}_a \tilde{\mathcal{O}}^{ij}_a,
\end{align}
where $i,j$ are quark-flavor indices. 
The effective operators are defined as\footnote{
Besides, there is a strong CP phase, $\bar\theta$. 
In this letter, we assume $\bar\theta = 0$, for simplicity. 
}
\begin{align}
\mathcal{O}_1^i &=-\frac{i}{2} m_{d_i} \bar{d}_i e Q_d (F\cdot \sigma) \gamma_5 d_i,
\\
\mathcal{O}_2^i &= -\frac{i}{2} m_{d_i} \bar{d}_i g_s (G\cdot \sigma) \gamma_5 d_i,
\\
\mathcal{O}_3 &= -\frac{1}{6} g_s f^{ABC} \epsilon^{\mu\nu\rho\sigma} G^A_{\mu\lambda} {G^B_{\nu}}^{\lambda} G^C_{\rho\sigma},
\\
\mathcal{O}_4^i &= (\bar{d}_i^{\alpha} {d}_i^{\alpha}) (\bar{d}_j^{\beta}i\gamma_5 {d}_j^{\beta}),
\\
\mathcal{O}_5^i &= (\bar{d}_i^{\alpha} \sigma^{\mu\nu} {d}_i^{\alpha}) (\bar{d}_j^{\beta} i\sigma_{\mu\nu}\gamma_5 {d}_j^{\beta}),
\\
\tilde{\mathcal{O}}^{ij}_1 &= (\bar{d}_i^{\alpha} {d}_i^{\alpha}) (\bar{d}_j^{\beta} i\gamma_5 {d}_j^{\beta}),
\\
\tilde{\mathcal{O}}^{ij}_2 &= (\bar{d}_i^{\alpha} {d}_i^{\beta}) (\bar{d}_j^{\beta} i\gamma_5 {d}_j^{\alpha}),
\\
\tilde{\mathcal{O}}^{ij}_3 &= (\bar{d}_i^{\alpha} \sigma^{\mu\nu} {d}_i^{\alpha}) (\bar{d}_j^{\beta} i\sigma_{\mu\nu} \gamma_5 {d}_j^{\beta}),
\\
\tilde{\mathcal{O}}^{ij}_4 &= (\bar{d}_i^{\alpha} \sigma^{\mu\nu} {d}_i^{\beta}) (\bar{d}_j^{\beta} i \sigma_{\mu\nu} \gamma_5 {d}_j^{\alpha}), 
\label{EDMeff}
\end{align}
where $\alpha, \beta$ are color indices, and $F_{\mu\nu}$ $(G^A_{\mu\nu})$ is the electromagnetic (gluon) field strength. 
We define $F\cdot \sigma = F_{\mu\nu} \sigma^{\mu\nu}$, $G\cdot \sigma = G^A_{\mu\nu}\sigma^{\mu\nu}T^A$ and $\tilde{G}^A_{\mu\nu} = \frac{1}{2}\epsilon_{\mu\nu\rho\sigma} {G^A}^{\rho\sigma}$ with $\sigma^{\mu\nu}=\frac{i}{2}[\gamma^{\mu},\gamma^{\nu}]$ and $\epsilon^{0123}=+1$. 
Also, $f^{ABC}$ is the structure constant, and $m_q$ is a mass for quark $q$. 
These operators are mixed through the RGEs, which are found in Refs.~\cite{Shifman:1976de,Dai:1989yh,Braaten:1990zt,Boyd:1990bx,Hisano:2012cc} (see Refs.~\cite{Degrassi:2005zd,Brod:2018pli,deVries:2019nsu} for higher order corrections).

The matching conditions are derived by integrating out SM heavy degrees of freedom, such as $W, Z, H$ and $t$.
At the tree level, we obtain the conditions,
\begin{align}
(\tilde{C}_1^{ij})^{\rm tree} &= \frac{i}{4} \Big[ (C^{(8)}_{qd})_{jiij} - (C^{(8)}_{qd})_{ijji} \Big],
\label{eq:C1tree}
\\
(\tilde{C}_2^{ij})^{\rm tree} &= \frac{i}{4} \Big[
 2\Big( (C^{(1)}_{qd})_{jiij}-(C^{(1)}_{qd})_{ijji} \Big) 
 -\frac{1}{N_c} \Big((C^{(8)}_{qd})_{jiij}-(C^{(8)}_{qd})_{ijji} \Big)
\Big],
\label{eq:C2tree} 
\end{align}
where the Wilson coefficients are evaluated at the EWSB scale, $\mu=\mu_{W}$.
The other LEFT operators are not induced at the tree level.

In addition, the SMEFT $\Delta F=1$ operators can generate $\Delta F=0$ amplitudes through the one-loop matching conditions at the EWSB scale.
We focus on the contributions from the loop diagrams with the top quark and the $W$ boson (cf.~Ref.~\cite{Endo:2018gdn}).
The conditions in the Feynman-'t Hooft gauge are obtained as
\begin{align}
(\tilde{C}_1^{ij})^{\rm 1\textrm{--}loop} =&
-\frac{\alpha}{2 \pi s_W^2} {\rm Im} \Big[ \lambda_t^{ji} (C^{(8)}_{ud})_{33ij} \Big] I_1(x_t,\mu_W)
-\frac{\alpha}{  \pi s_W^2} {\rm Im} \Big[ \lambda_t^{ji} (C^{(8)}_{qd})_{33ij} \Big] J(x_t)
\notag \\ &
+\frac{\alpha}{4\pi s_W^2} \sum_{m=1}^3 \bigg\{
 {\rm Im} \Big[ \lambda_t^{jm} (C^{(8)}_{qd})_{miij} \Big]
+{\rm Im} \Big[ \lambda_t^{mi} (C^{(8)}_{qd})_{jmij} \Big] 
\bigg\} K (x_t,\mu_W),
\label{eq:edm1mat}
\\
%%%
(\tilde{C}_2^{ij})^{\rm 1\textrm{--}loop} =&
-\frac{\alpha}{\pi s_W^2} {\rm Im} \Big\{
 \lambda_t^{ji} \Big[ (C^{(1)}_{ud})_{33ij}-\frac{1}{2 N_c} (C^{(8)}_{ud})_{33ij}-(C_{Hd})_{ij} \Big]
\Big\} I_1(x_t,\mu_W)
\notag \\ &
-\frac{2\alpha}{\pi s_W^2} {\rm Im} \Big\{
 \lambda_t^{ji} \Big[ (C^{(1)}_{qd})_{33ij} -\frac{1}{2 N_c}(C^{(8)}_{qd})_{33ij} \Big] \Big\} J(x_t)
\notag \\ &
+ \frac{\alpha}{2\pi s_W^2} \sum_{m=1}^3 \bigg\{
{\rm Im} \Big[ 
  \lambda_t^{jm} \Big[ (C^{(1)}_{qd})_{miij}-\frac{1}{2N_c} (C^{(8)}_{qd})_{miij} \Big] 
\Big]
\notag \\ &
+ {\rm Im} \Big[ 
  \lambda_t^{mi} \Big[ (C^{(1)}_{qd})_{jmij}-\frac{1}{2N_c} (C^{(8)}_{qd})_{jmij} \Big] 
\Big]
\bigg\} K (x_t,\mu_W),
\label{eq:edm2mat}
\end{align}
where the parameters are defined as
\begin{align}
 x_t = \frac{m_t^2}{M_W^2},~~~
 \lambda_t^{ij} = V_{ti}^{\ast} V_{tj}.
\end{align}
Here, $V_{ij}$ is the CKM matrix, and $s_W = \sin\theta_W$ with the Weinberg angle $\theta_W$.
The loop functions are defined as
\begin{align}
 I_1(x,\mu) &=
 \frac{x}{8} \left[ 
 \ln\frac{\mu}{M_W} - \frac{x-7}{4(x-1)} - \frac{x^2-2x+4}{2(x-1)^2} \ln x 
 \right],
\\
 J(x) &= 
 \frac{x}{16} \left( 1 - \frac{2\ln x}{x-1} \right),
 \\
 K(x,\mu) &= 
 \frac{x}{8} \left[
 \ln \frac{\mu}{M_W} + \frac{3(x+1)}{4(x-1)} - \frac{x(x+2)}{2(x-1)^2} \ln x
 \right].
\end{align}
All the Wilson coefficients are evaluated at the EWSB scale, $\mu=\mu_{W}$.
The other LEFT operators for the EDMs do not receive one-loop corrections at this scale.  

The SMEFT $\Delta F=1$ operators also generate LEFT $\Delta F=2$ operators.
The latter operators are defined as
\begin{align}
\mathcal{H}_{\rm eff}^{\Delta F=2}&= (C_1)_{ij} (\bar{d}_i \gamma^{\mu} P_L d_j) (\bar{d}_i \gamma_{\mu}P_L d_j)\notag
\\
& +(C_2)_{ij} (\bar{d}_i P_L d_j) (\bar{d}_i P_L d_j) +(C_3)_{ij} (\bar{d}^{\alpha}_i P_L d^{\beta}_j) (\bar{d}^{\beta}_i P_L d^{\alpha}_j)\notag
\\
&+(C_4)_{ij} (\bar{d}_i P_L d_j) (\bar{d}_i P_R d_j) +(C_5)_{ij} (\bar{d}^{\alpha}_i P_L d^{\beta}_j) (\bar{d}^{\beta}_i P_R d^{\alpha}_j)\notag
\\
&+(C'_1)_{ij} (\bar{d}_i \gamma^{\mu} P_R d_j) (\bar{d}_i \gamma_{\mu} P_R d_j)\notag
\\
&+(C'_2)_{ij} (\bar{d}_i P_R d_j) (\bar{d}_i P_R d_j) +(C'_3)_{ij} (\bar{d}^{\alpha}_i P_R d^{\beta}_j) (\bar{d}^{\beta}_i P_R d^{\alpha}_j).
\end{align}
We follow the analysis in Ref.~\cite{Endo:2018gdn}, where the SMEFT RGEs and the matching formulae at the one-loop level are provided.

Below the EWSB scale, the LEFT $\Delta F=0,2$ operators are evolved by the RGEs.
Then, the low-scale observables are evaluated around the hadron scale.

%%%%%%%%%%%%%%%%%%%%%%%%%%%%%%%%%%%%%%%%%%%%%%%%%%%%%%%%
\section{Observables}
%%%%%%%%%%%%%%%%%%%%%%%%%%%%%%%%%%%%%%%%%%%%%%%%%%%%%%%%

In this section, low-scale observables are summarized.
We consider the EDMs, $\epsilon_K$ and $\Delta M_{B_d}$.
All of them are very sensitive to NP contributions to $CP$ violations.

%%%%%%%%%%%%%%%%%%%%%%%%%%%%%%%%%%%%%%%%%%%%%%%%%%%%%%%%
\subsection{Nuclear EDMs}

The $CP$-violating operators of the down-type quarks induce the nuclear EDMs.\footnote{
In the analysis, $CP$-violating baryon-meson interactions are considered to discuss the nuclear EDMs (see Appendix~\ref{sec:EDMchi}).
They can also induce the electron EDM, e.g., via the Barr-Zee diagram, which will be explored in future. 
}
Then, hadronic matrix elements are necessary to evaluate their contributions. 
There are many types of the SMEFT four-quark operators.
Contributions of $\tilde{\mathcal{O}}_1^{ds}$ and $\tilde{\mathcal{O}}^{sd}_1$ are evaluated by the effective chiral Lagrangian technique~\cite{deVries:2012ab}.
Those operators generate $CP$-violating baryon-meson interactions through vacuum-expectation values (VEVs) of pseudoscalar mesons.
Then, the $^{199}$Hg EDM is induced at the tree level as~\cite{Bijnens:1996np}
\begin{align}
\frac{d_{\rm Hg}}{e}\sim \left(0.005\tilde{C}_1^{ds} -0.032 \tilde{C}_1^{sd}  \right){\rm GeV^{-1}}.\label{eq:HgEDMds}
\end{align}
In addition, from the baryon-meson loop diagrams, we obtain
\footnote{The nucleon EDMs are also induced by baryon-meson diagrams at the tree level~\cite{Bijnens:1996np}. However, we confirmed that they are sub-dominant.}
\begin{align}
\frac{d_n}{e} &\sim \left( -0.026 \tilde{C}_1^{ds} + 0.169 \tilde{C}_1^{sd}\right){\rm GeV^{-1}},\label{eq:nEDMds}
\\
\frac{d_p}{e} &\sim \left(  0.023 \tilde{C}_1^{ds} - 0.149 \tilde{C}_1^{sd}\right){\rm GeV^{-1}},
\label{eq:pEDMds}
\end{align}
where the Wilson coefficients are estimated at the hadron scale, $\mu=1\,{\rm GeV}$.
Here and hereafter, we set $\bar{\theta}=0$ for simplicity\footnote{
The Peccei-Quinn (PQ) mechanism is not assumed for realizing $\bar{\theta}=0$. 
It is straightforward to extend the case for $\bar{\theta} \neq 0$. 
Then, the PQ mechanism is introduced to avoid the strong $CP$ problem. 
The following conclusions do not change qualitatively. }.
The derivations of Eqs.~\eqref{eq:nEDMds} and \eqref{eq:pEDMds} are given in Appendix~\ref{sec:EDMchi}.

Four-quark operators, $\tilde{\mathcal{O}}^{db}$ and $\tilde{\mathcal{O}}^{bd}$, involve the bottom quark. 
In order to derive their contributions to the neutron and proton EDMs, we follow the strategy explored in Refs.~\cite{Demir:2002gg,Demir:2003js,Haisch:2019bml}.
The result becomes
\begin{align}
\frac{d_n}{e} \sim 4.2 \times 10^{-4} 
\left( \tilde{C}^{bd} + 0.75 \tilde{C}^{db} \right){\rm GeV^{-1}},\label{eq:nEDMdb}
\\
\frac{d_p}{e} \sim 6.1 \times 10^{-4} 
\left( \tilde{C}^{bd} + 0.75 \tilde{C}^{db} \right){\rm GeV^{-1}},\label{eq:pEDMdb}
\end{align}
where the Wilson coefficients are estimated at the hadron scale, $\mu=1\,{\rm GeV}$.
Here, the contribution to the proton EDM, \eqref{eq:pEDMdb}, is derived by multiplying a ratio of the magnetic moments, $\mu_p/\mu_n$, to that of the neutron EDM, \eqref{eq:nEDMdb} (cf., Ref.~\cite{Haisch:2019bml}).
On the other hand, $\tilde{\mathcal{O}}^{sb}$ and $\tilde{\mathcal{O}}^{bs}$ are much less constrained by the EDMs, because they do not depend on the down quark.

Let us summarize the current experimental limits and future prospects. 
The current bounds are obtained as~\cite{Griffith:2009zz,Graner:2016ses,Afach:2015sja,Sahoo:2016zvr}
\begin{align}
|d_{\rm Hg}| & < 6.3 \times 10^{-30}~e\, {\rm cm},~~~[90 \%~{\rm C.L.}] \\
|d_n| &< 3.0 \times 10^{-26}~e\, {\rm cm},~~~[90 \%~{\rm C.L.}] \\
|d_p| &< 2.1 \times 10^{-25}~e\, {\rm cm}.
\end{align}
In future, experiments are projected to achieve the sensitivities of $|d_n| \sim 10^{-28}~e\,{\rm cm}$\cite{Strategy:2019vxc} and $|d_p| \sim 10^{-29}~e\,{\rm cm}$~\cite{Anastassopoulos:2015ura}.
Although the $^{199}$Hg EDM constraint is the strongest at this moment, the neutron/proton EDMs can provide severer bound by the future experiments.

Before closing this section, let us comment on contributions to the electric and chromoelectric dipole moments, $\mathcal{O}^{i}_1$ and $\mathcal{O}^{i}_2$. 
As mentioned in the previous section, the SMEFT $\Delta F=1$ operators contribute only to the LEFT four-quark operators, $\tilde{\mathcal{O}}^{ij}_{1,2}$. 
Below the EWSB scale, they induce $\mathcal{O}^{i}_{1,2}$ through radiative corrections.
However, according to the RGEs in the LEFT, their contributions appear as linear combinations of $\tilde{C}^{ij}_1 +\tilde{C}^{ji}_1$ and $\tilde{C}^{ij}_2 +\tilde{C}^{ji}_2$ as
\begin{align}
C_{a}^i = \alpha_{a} (\tilde{C}^{ij}_1 +\tilde{C}^{ji}_1) + \beta_{a} (\tilde{C}^{ij}_2 +\tilde{C}^{ji}_2),
\end{align}
for $\mathcal{O}^{i}_{1,2}$ with $a=1,2$ and coefficients $\alpha_a,\beta_a$.
By substituting the SMEFT contributions into $\tilde{C}^{ij}_{1,2}$ in the right-hand side, all the contributions are found to vanish (see Eqs.~\eqref{eq:C1tree}--\eqref{eq:edm2mat}).
Consequently, the SMEFT $\Delta F=1$ operators do not generate the electric or chromoelectric dipole moment.
Hence, we will study the nuclear EDMs directly from the four-quark operators.

%%%%%%%%%%%%%%%%%%%%%%%%%%%%%%%%%%%%%%%%%%%%%%%%%%%%%%%%
\subsection{$\Delta F=2$ observables}

The $\Delta F=2$ operators contribute to the oscillations of the neutral mesons.
In particular, the indirect $CP$ violation of the neutral $K$ mesons, $\epsilon_K$, and the mass difference of the neutral $B_q$ mesons are sensitive to NP contributions. 
The former is sensitive to flavor violations between the first two generations of the down-type quark. 
The SM and NP contributions are represented as
\begin{align}
\epsilon_K =e^{i\phi_{\epsilon}} \left(\epsilon_K^{\rm SM} +\epsilon_K^{\rm NP} \right),
\end{align}
with $\phi_{\epsilon}=(43.51\pm 0.05)^{\circ}$.
The SM prediction is estimated as~\cite{Bailey:2018feb}
\begin{align}
\epsilon_K^{\rm SM}= (2.035 \pm 0.178)\times 10^{-3},
\label{eq:epSM}
\end{align}
where $V_{cb}$ is determined by the inclusive semileptonic $B$ decays.
The NP contribution is represented as
\begin{align}
\epsilon_K^{\rm NP} =\frac{\tilde{\kappa}_{\epsilon}}{\sqrt{2}(\Delta M_K)_{\rm exp}}\left[{\rm Im}\,(M_{12}^K)^{\rm NP} \right],
\end{align}
where $\tilde{\kappa}_{\epsilon} =0.94$~\cite{Buras:2008nn,Buras:2010pza} and $(\Delta M_K)_{\rm exp} = 3.483 \times 10^{-15}~{\rm GeV}$~\cite{Tanabashi:2018oca} are used. Also, $M_{12}^K={\langle K^0|\mathcal{H}^{\Delta S=2}_{\rm eff}|\bar{K}^0 \rangle}/2M_K$ with $M_K=0.4976~{\rm GeV}$~\cite{Tanabashi:2018oca}. The Wilson coefficients are evaluated with the NLO-QCD RGEs~\cite{Buras:2001ra}, and hadron matrix elements in Ref.~\cite{Garron:2016mva} are used.
On the other hand, the experimental result is~\cite{Tanabashi:2018oca}
\begin{align}
|\epsilon_K^{\rm exp}|= (2.228\pm 0.011)\times 10^{-3}.
\label{eq:epEX}
\end{align}
From Eqs.~\eqref{eq:epSM} and \eqref{eq:epEX}, we obtain the bound on the NP contribution as
\begin{align}
-0.16 \times 10^{-3} < \epsilon_K^{\rm NP} <0.55 \times 10^{-3},
\end{align}
at the $2\sigma$ level.

Next, flavor violations including the bottom quark are constrained by the oscillations of the neutral $B_q$ mesons. 
In particular, those between the first and third generations contribute to $\Delta M_{B_d}$.
The SM and NP contributions are represented as
\begin{align}
\Delta M_{B_d} = 2~\left|(M_{12}^{B_d})^{\rm SM}+(M_{12}^{B_d})^{\rm NP}\right| 
\equiv \Delta M_{B_d}^{\rm SM} +\Delta M_{B_d}^{\rm NP},
\end{align}
where $M_{12}^{B_d} = {\langle B^0|\mathcal{H}^{\Delta B=2}_{\rm eff}|\bar{B}^0 \rangle}/2M_{B_d}$ with $M_{B_d} = 5.27958~{\rm GeV}$~\cite{Tanabashi:2018oca}.
The first term in the right-hand side denotes the SM contribution, which is estimated as~\cite{Bazavov:2016nty}
\begin{align}
\Delta M_d^{\rm SM}=(4.21 \pm 0.34)\times 10^{-13}\,{\rm GeV}.
\end{align}
The Wilson coefficients are evaluated with the NLO-QCD RGEs~\cite{Buras:2001ra}, and hadron matrix elements in Ref.~\cite{Bazavov:2016nty} are used.
On the other hand, the experimental result is obtained as~\cite{Tanabashi:2018oca}
\begin{align}
\Delta M_d^{\rm exp}= (3.3338\pm 0.0125)\times 10^{-13}\,{\rm GeV}.
\end{align}
Thus, the NP contribution is required to satisfy,
\begin{align}
0.20\times 10^{-13} < \Delta M_d^{\rm NP} <1.56 \times 10^{-13},
\end{align}
at the $2\sigma$ level.
Finally, although $\Delta M_{B_s}$ gives a constraint on flavor violations between the second and third generations, the bound from the EDMs is very weak (see Eq.~\eqref{eq:nEDMdb}). 
Hence, we do not consider them in this letter.

%%%%%%%%%%%%%%%%%%%%%%%%%%%%%%%%%%%%%%%%%%%%%%%%%%%%%%%%
\section{Numerical analysis}
%%%%%%%%%%%%%%%%%%%%%%%%%%%%%%%%%%%%%%%%%%%%%%%%%%%%%%%%

\begin{figure}[th]
\begin{center}
\includegraphics[scale=0.65, bb= 100 0 290 161]{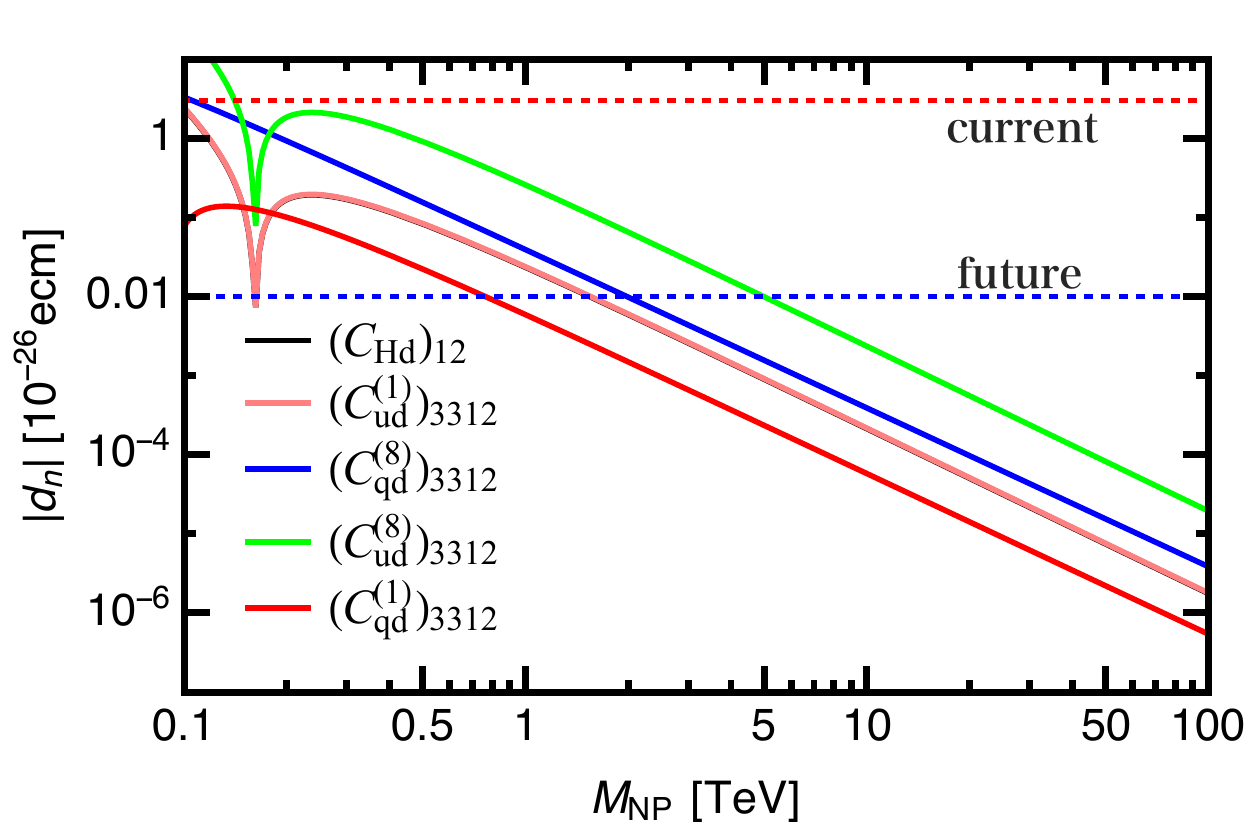}\hspace{25mm}
\includegraphics[scale=0.65, bb= 0 0 256 160]{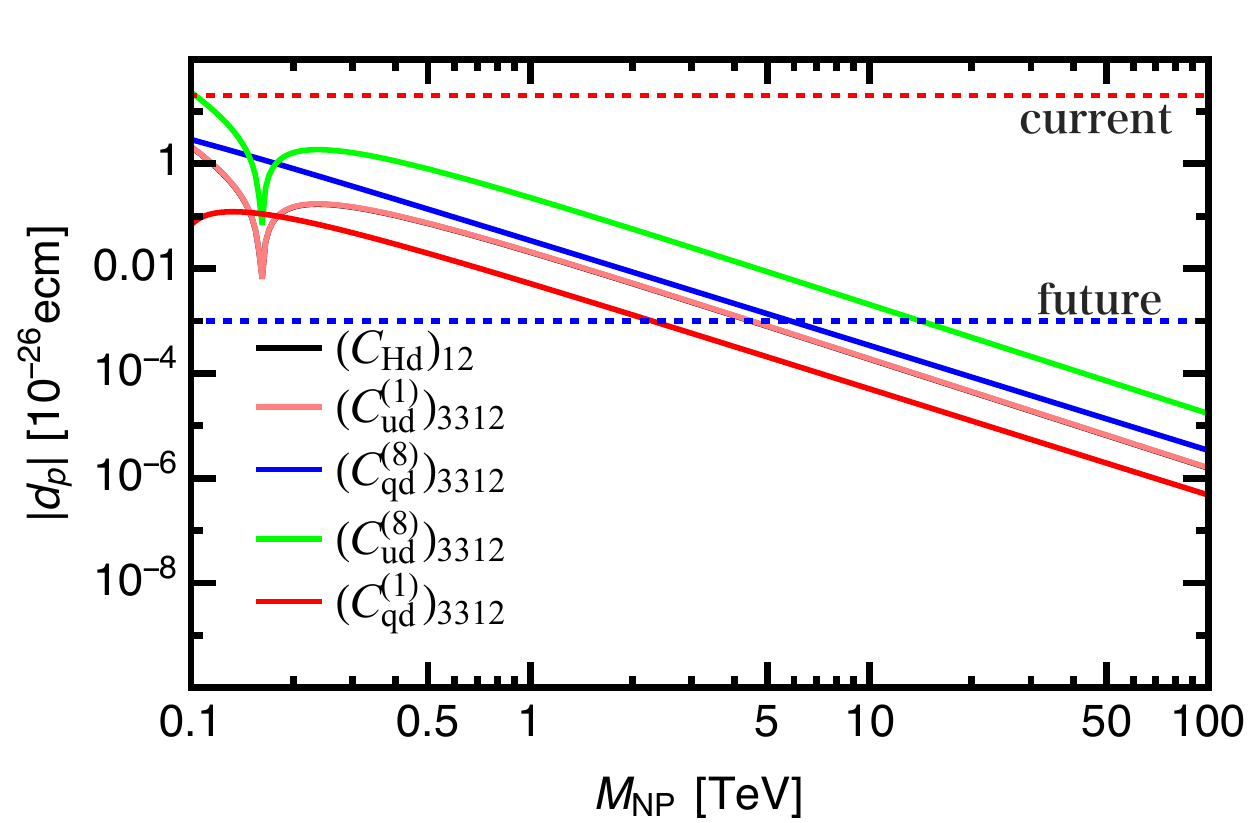}
\end{center}
\begin{center}
\includegraphics[scale=0.65, bb= 100 0 290 230]{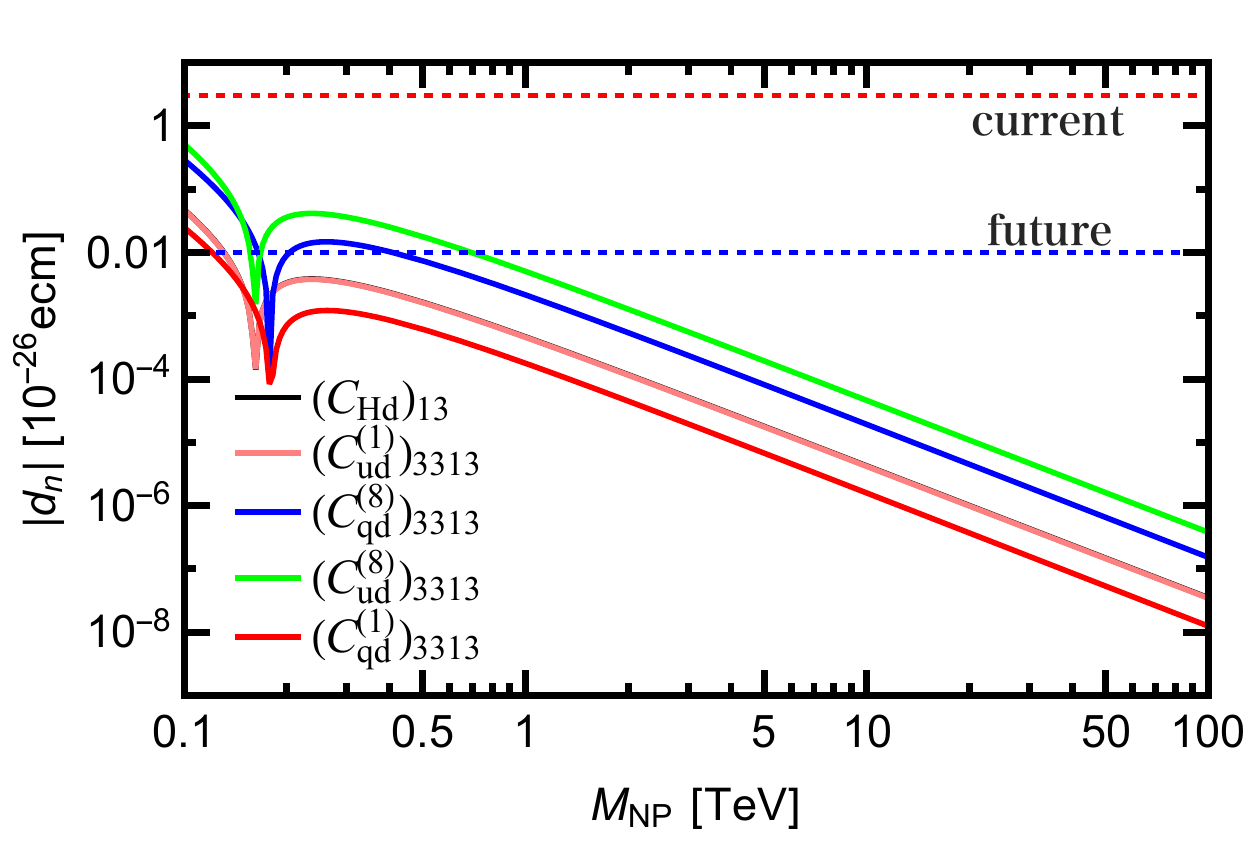}\hspace{25mm}
\includegraphics[scale=0.65, bb= 0 0 256 160]{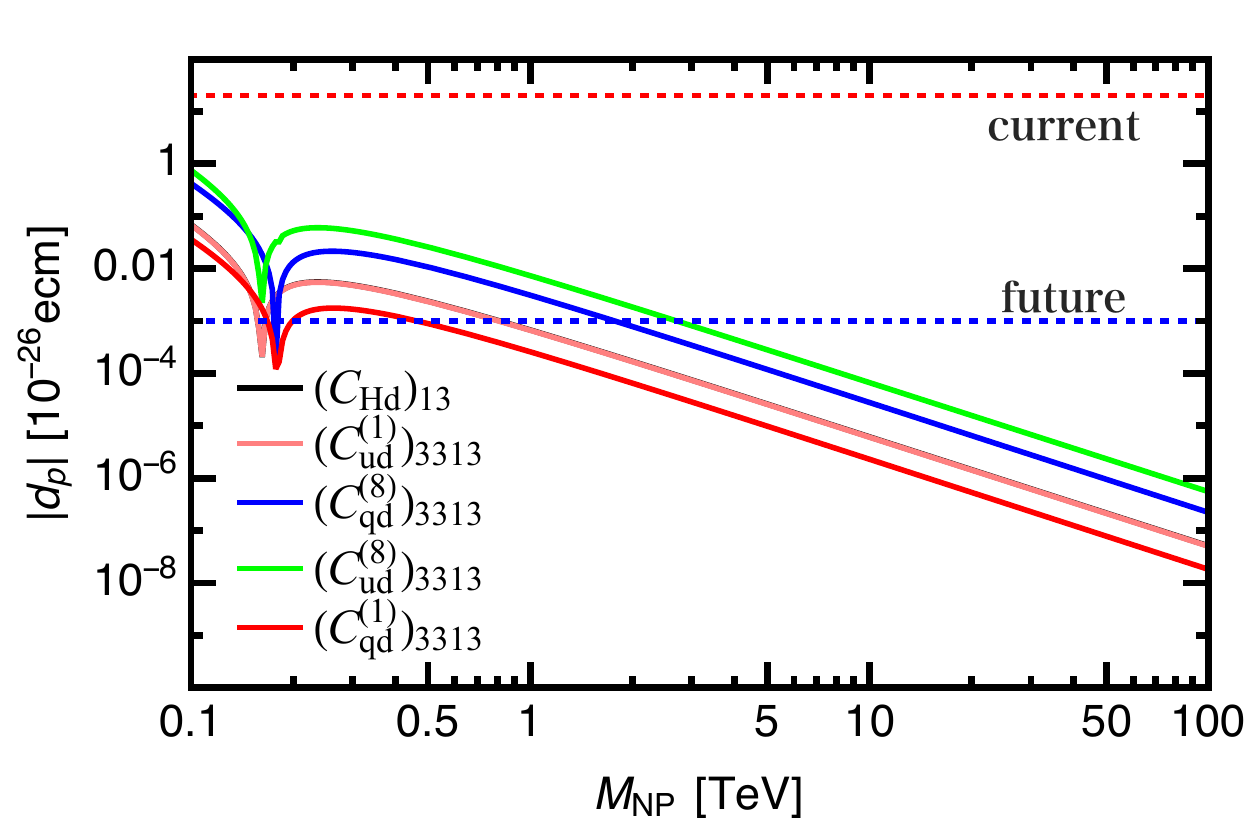}
\end{center}
\includegraphics[scale=0.65, bb= -190 0 200 230]{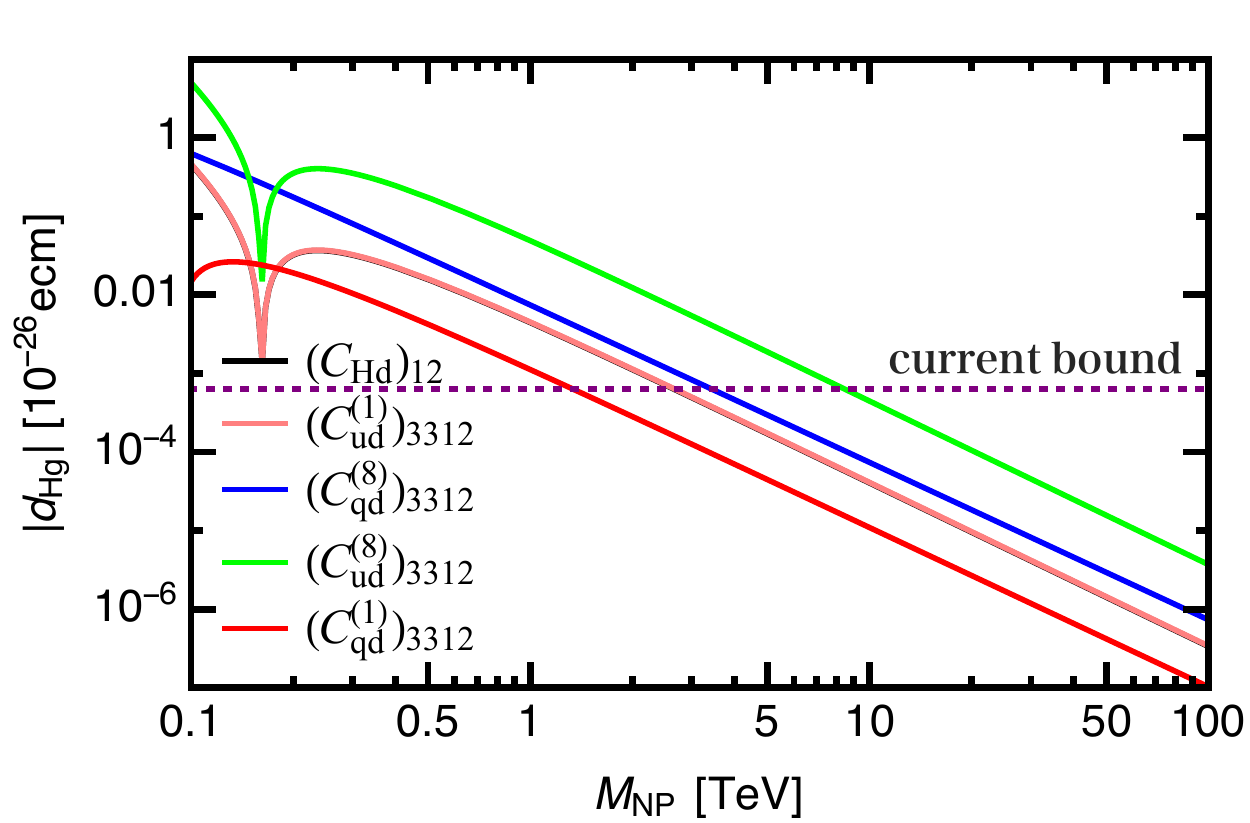}
\caption{
The neutron and proton EDMs are estimated for the SMEFT $\Delta F=1$ operators of the down and strange quarks with the top quarks in the top panels. 
Those for the down and bottom quarks are shown in the middle panels.
The red and blue dotted lines are the current experimental limit and the future sensitivity.
Also, the $^{199}$Hg EDM for the down and strange quarks are in the bottom panel.
The purple dotted line is the current experimental limit.
The Wilson coefficients are $i/M_{\rm NP}^{2}$ at the NP scale $M_{\rm NP}$. 
}
\label{fig:np_ver}
\end{figure}

In this section, we study contributions of the SMEFT $\Delta F=1$ operators to the nuclear EDMs and $\Delta F=2$ observables, $\epsilon_K$ and $\Delta M_{B_d}$. 
In Fig.~\ref{fig:np_ver}, the neutron, proton and $^{199}$Hg EDMs are estimated.
On each line, one of the Wilson coefficients is set to be $C_i = i/M_{\rm NP}^{2}$ at the NP scale, $M_{\rm NP}$. 
The other coefficients are zero. 
The effective operators missing in the list do not contribute to the EDMs as well as the $\Delta F=2$ observables.\footnote{
There are operators which can contribute to the EDMs through self-energy corrections. 
The matching conditions are provided in Section \ref{sec:formula}, and it is straightforward to analyze them.}
Once the operator is set, the RGEs are solved, and the matching conditions are taken into account.
In the top and bottom panels, the four-quark operators mix the first two generations of the down-type quark.
On the other hand, the operators in the middle panels include the bottom quark. 
In low $M_{\rm NP}$ regions, it is found that the EDMs are suppressed, where the loop functions, $I_1$ and $K$, vanish.

For the plots of the neutron and proton EDMs, the horizontal red and blue dotted lines correspond to the current experimental bound and the future sensitivity, respectively. 
For the latter, we quote $|d_n| = 10^{-28}~e\,{\rm cm}$ and $|d_p| = 10^{-29}\,e\,{\rm cm}$.
Currently, the neutron EDM excludes $M_{\rm NP} \lesssim 100\,{\rm GeV}$ of $(C^{(1,8)}_{ud})_{3312}$ and $(C^{(1)}_{qd})_{3312}$.
On the other hand, the severest constraint is provided by the $^{199}$Hg EDM; the current experimental bound is shown by the horizontal purple dotted line in the bottom plot of Fig.~\ref{fig:np_ver}.
It is found that the NP contributions have already been excluded for $M_{\rm NP} \lesssim 1\text{--}9\,{\rm TeV}$.
The sensitivities of the neutron/proton EDMs are expected to be improved greatly.
They can probe the NP scale up to $2\text{--}10\,{\rm TeV}$, which are beyond the limit of the $^{199}$Hg EDM. 

The contributions to the nuclear EDMs are suppressed for the operators including the bottom quark. 
This is because the hadron matrix elements of such operators are small (see Eq.~\eqref{eq:nEDMdb}).
Currently, the constraint is weaker than $M_{\rm NP} \lesssim 100\,{\rm GeV}$ according to the middle panels of the figure, and the sensitivity may reach at most 3\,TeV in future. 

%%%%%%%%%%%%%%%%%%%%%%%%%%%%%%%%%%%%%%%%%%%%
%%%%%%%%%%%%%%%%%%%%%%%%%%%%%%%%%%%%%%%%%%%%

%%%%%%%%%%%%%%%%%%%%%%%%%%%%%%%%%%%%%%%%%%%%
\begin{figure}[t]
\begin{center}
\includegraphics[scale=0.65, bb= 110 0 256 254]{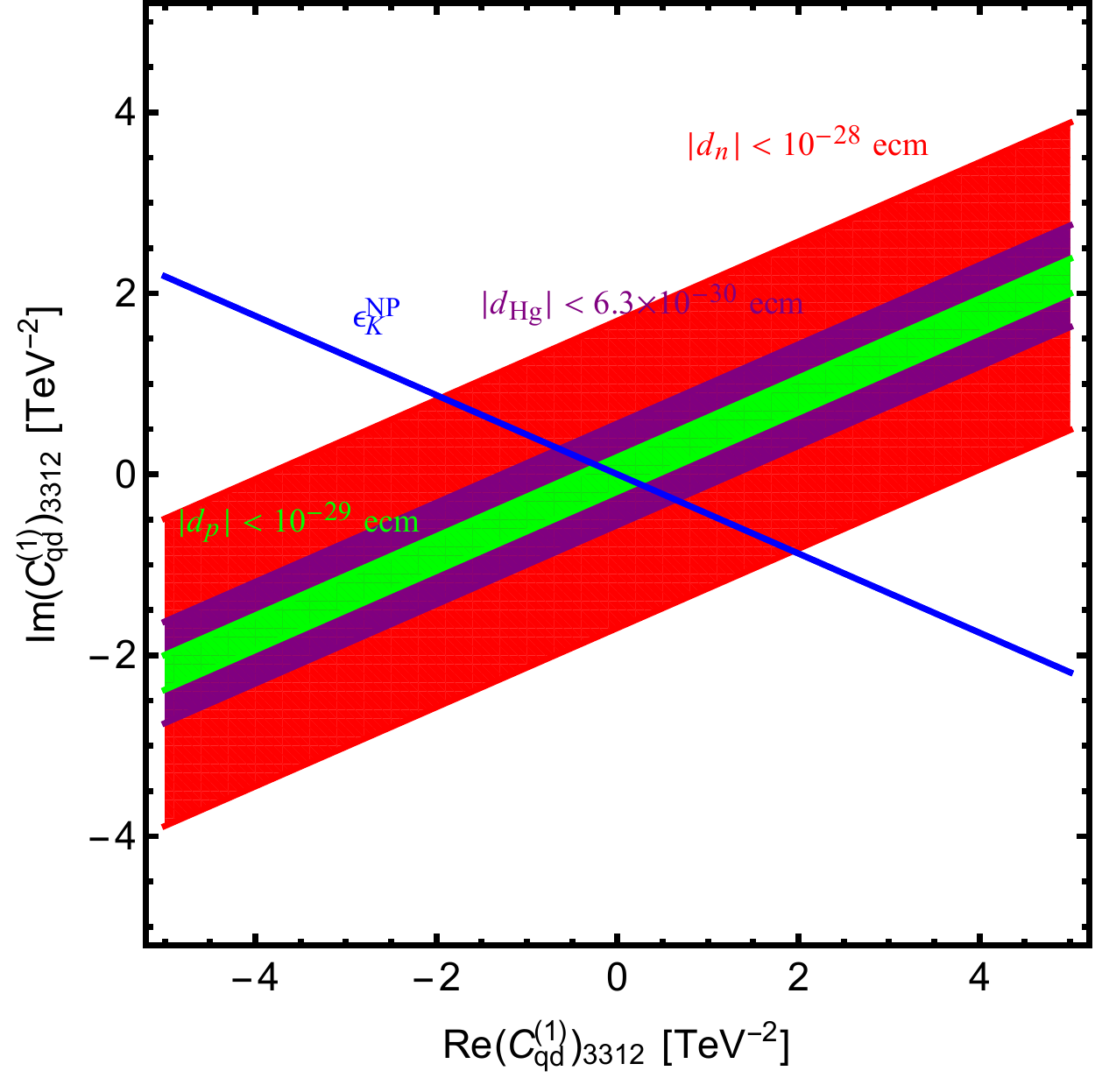}\hspace{25mm}
\includegraphics[scale=0.65, bb= 0 0 256 254]{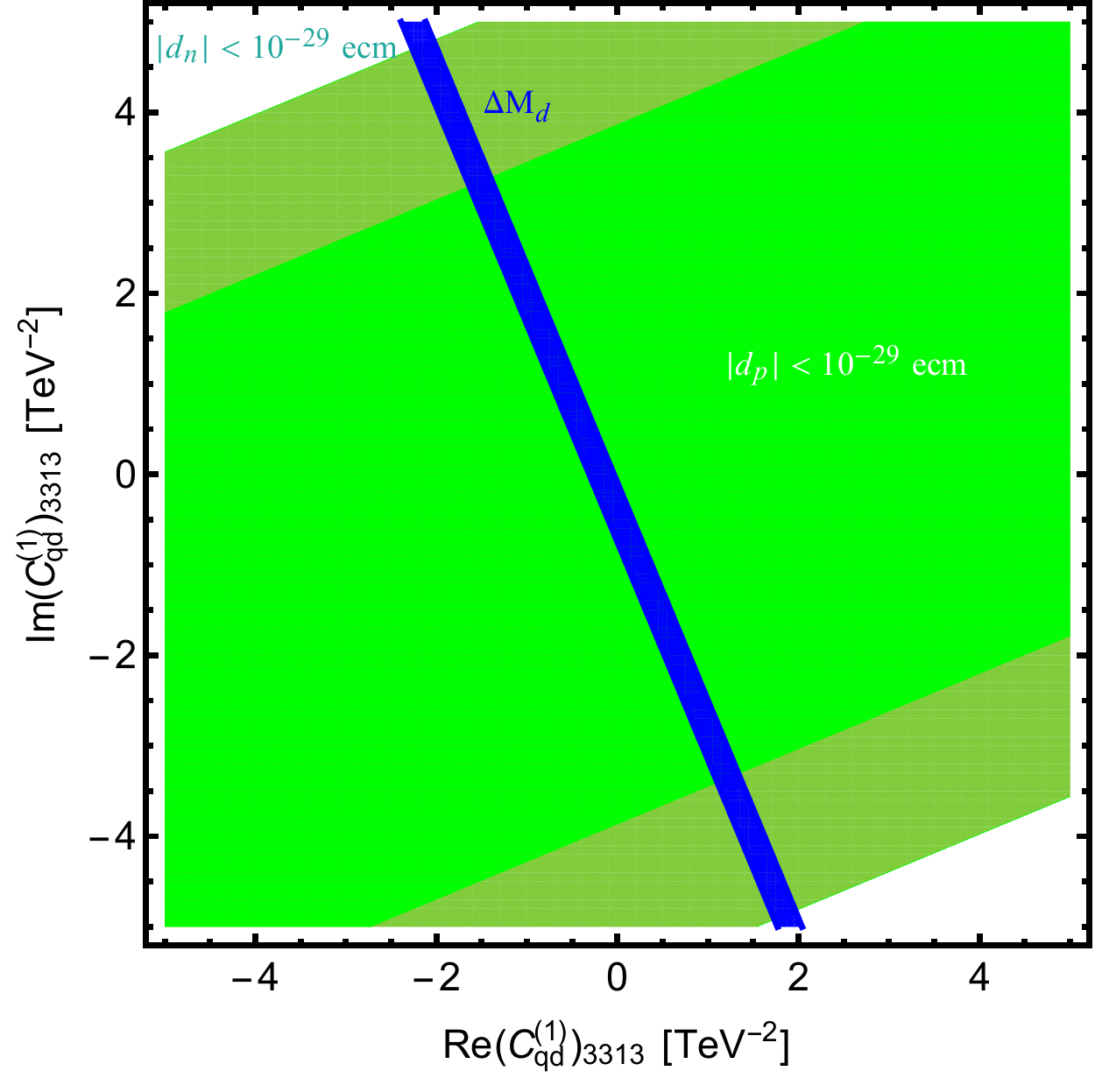}
\end{center}
\caption{
Contours of the EDMs and $\epsilon_K$ (left) and $\Delta M_{B_d}$ (right). 
Outside regions of the purple band are excluded by the $^{199}$Hg EDM, and those of the red and light green bands are probed by the future experiment in the left panel.
On the other hand, the deep green region in the right panel corresponds to $|d_n| < 10^{-29}~e\,{\rm cm}$, which is below the future sensitivity. 
In the left panel, the blue region is allowed by $\epsilon_K$ at the $2 \sigma$ level, and the region in the right panel is allowed by $\Delta M_{B_d}$ at the $2 \sigma$ level.
}
\label{fig:EDM_f_C1qd}
\end{figure}
%%%%%%%%%%%%%%%%%%%%%%%%%%%%%%%%%%%%%%%%%%%%
%%%%%%%%%%%%%%%%%%%%%%%%%%%%%%%%%%%%%%%%%%%%
\begin{figure}[t]
\begin{center}
\includegraphics[scale=0.65, bb= 110 0 256 254]{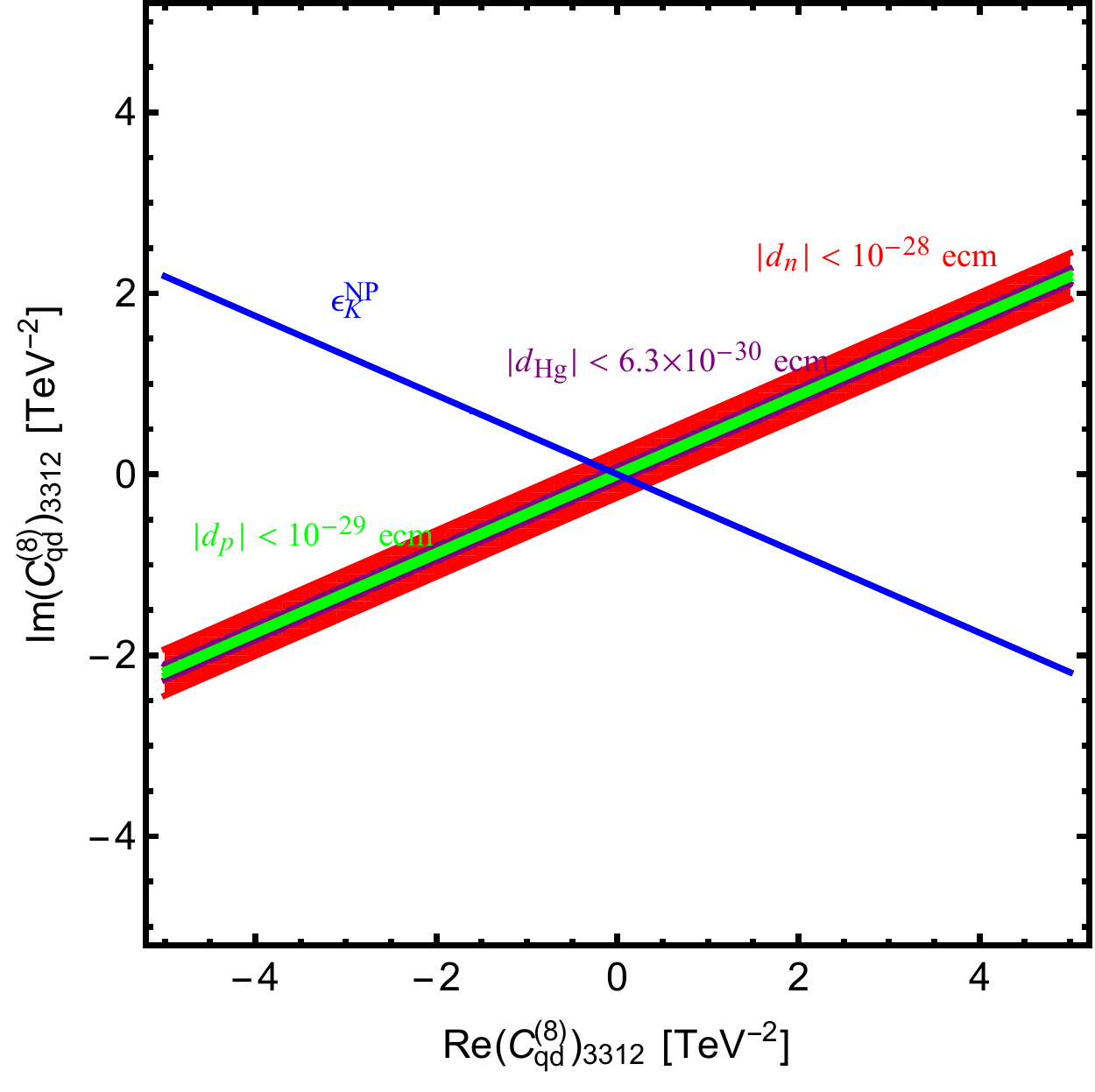}\hspace{25mm}
\includegraphics[scale=0.65, bb= 0 0 256 254]{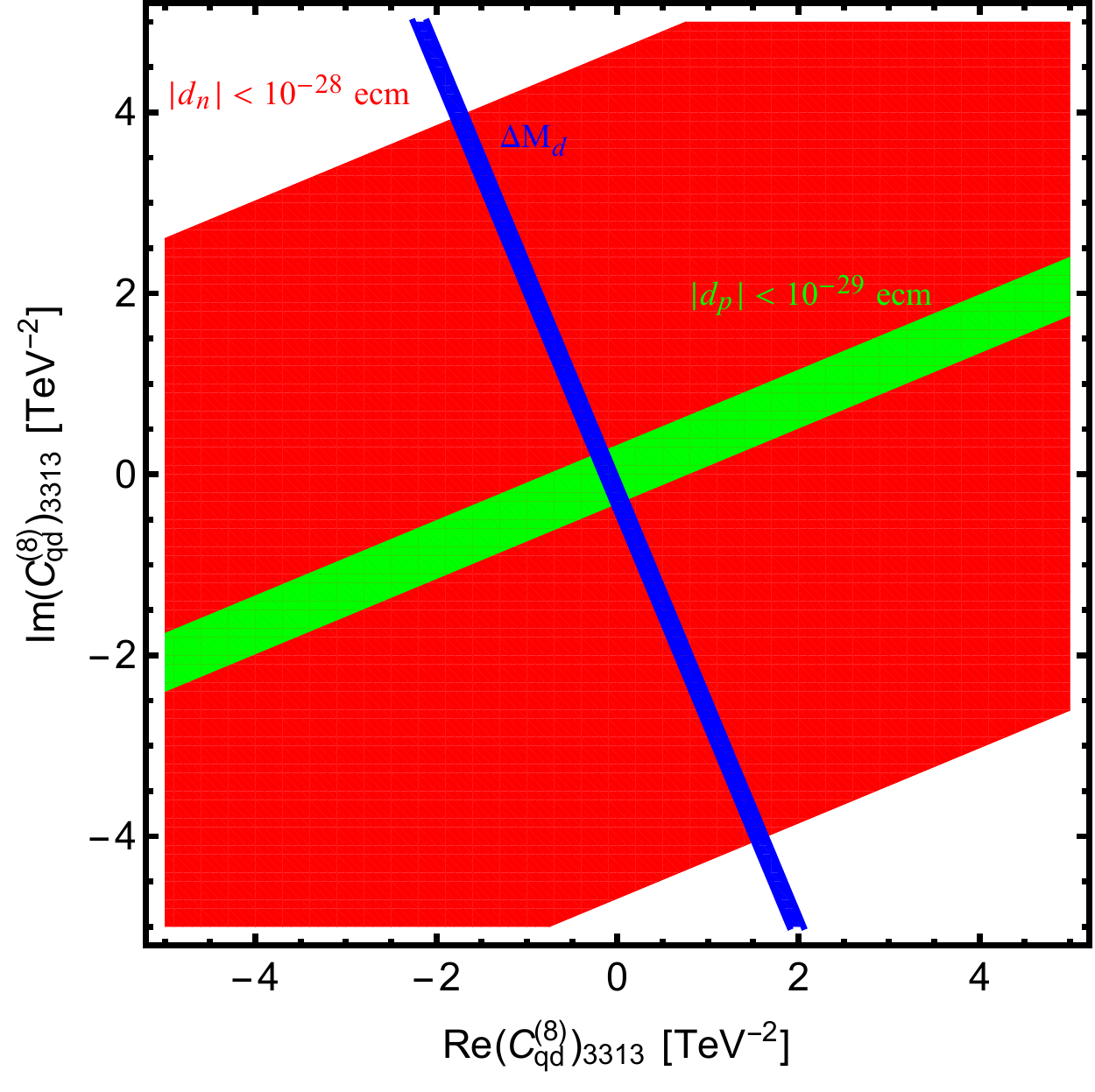}
\end{center}
\caption{
Same as Fig.~\ref{fig:EDM_f_C1qd}.
}
\label{fig:EDM_f_C8qd}
\end{figure}
%%%%%%%%%%%%%%%%%%%%%%%%%%%%%%%%%%%%%%%%%%%%
%%%%%%%%%%%%%%%%%%%%%%%%%%%%%%%%%%%%%%%%%%%%
\begin{figure}[h]
\begin{center}
\includegraphics[scale=0.65, bb= 110 0 253 252]{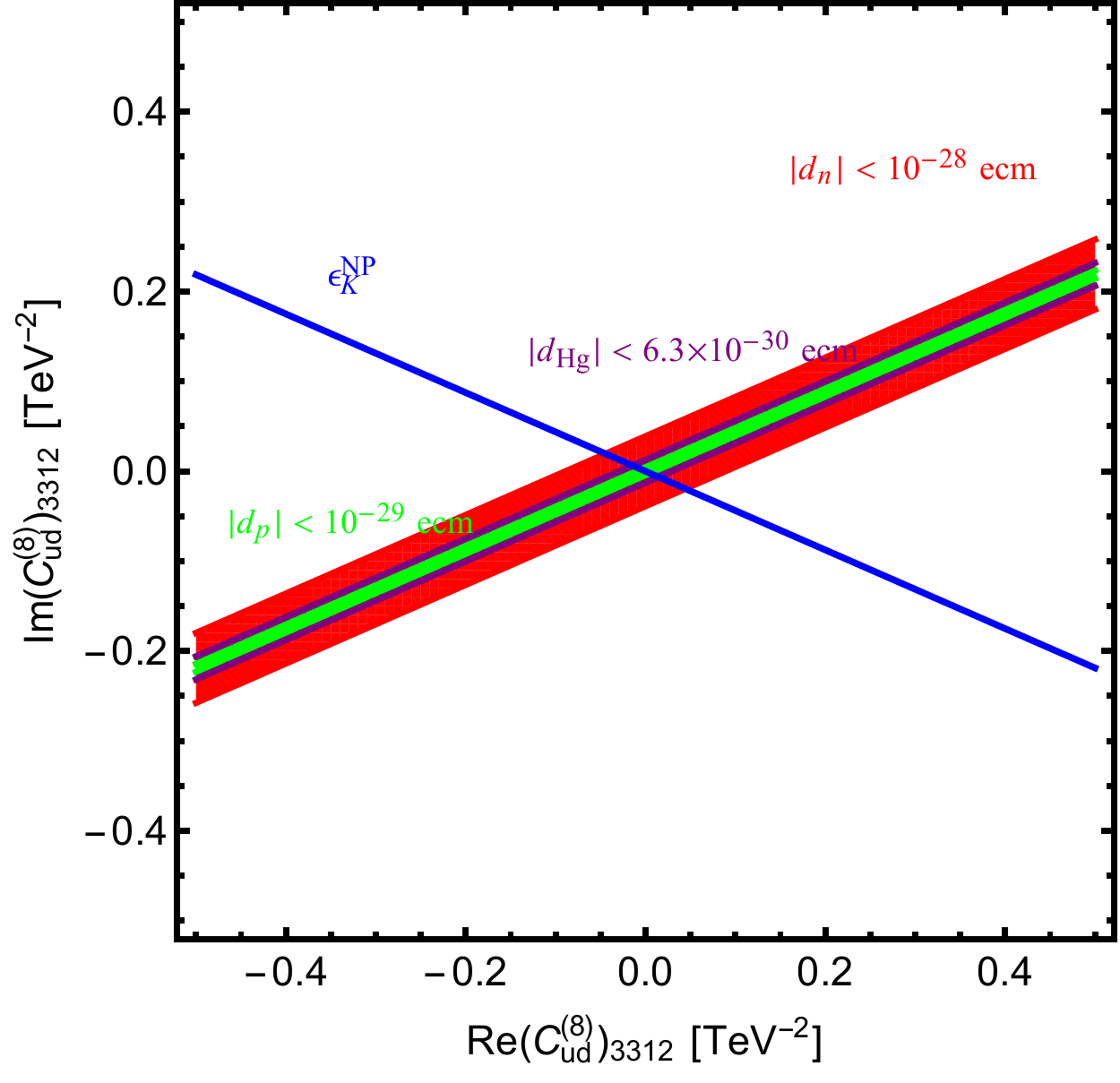}\hspace{25mm}
\includegraphics[scale=0.65, bb= 0 0 254 253]{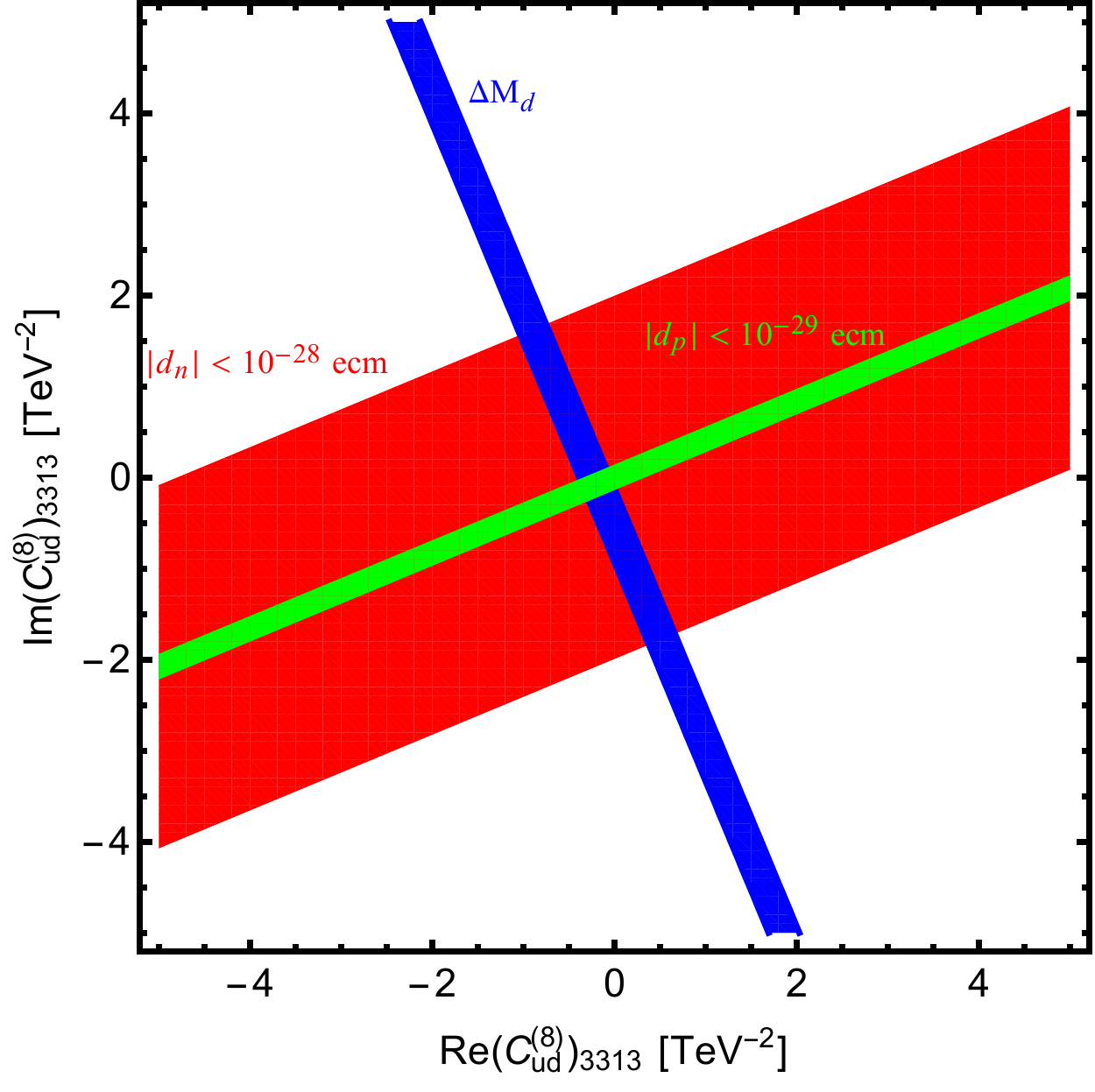}
\end{center}
\caption{
Same as Fig.~\ref{fig:EDM_f_C1qd}.
}
\label{fig:EDM_f_C8ud}
\end{figure}
%%%%%%%%%%%%%%%%%%%%%%%%%%%%%%%%%%%%%%%%%%%%
%%%%%%%%%%%%%%%%%%%%%%%%%%%%%%%%%%%%%%%%%%%%
\begin{figure}[t]
\begin{center}
\includegraphics[scale=0.65, bb= 110 0 256 256]{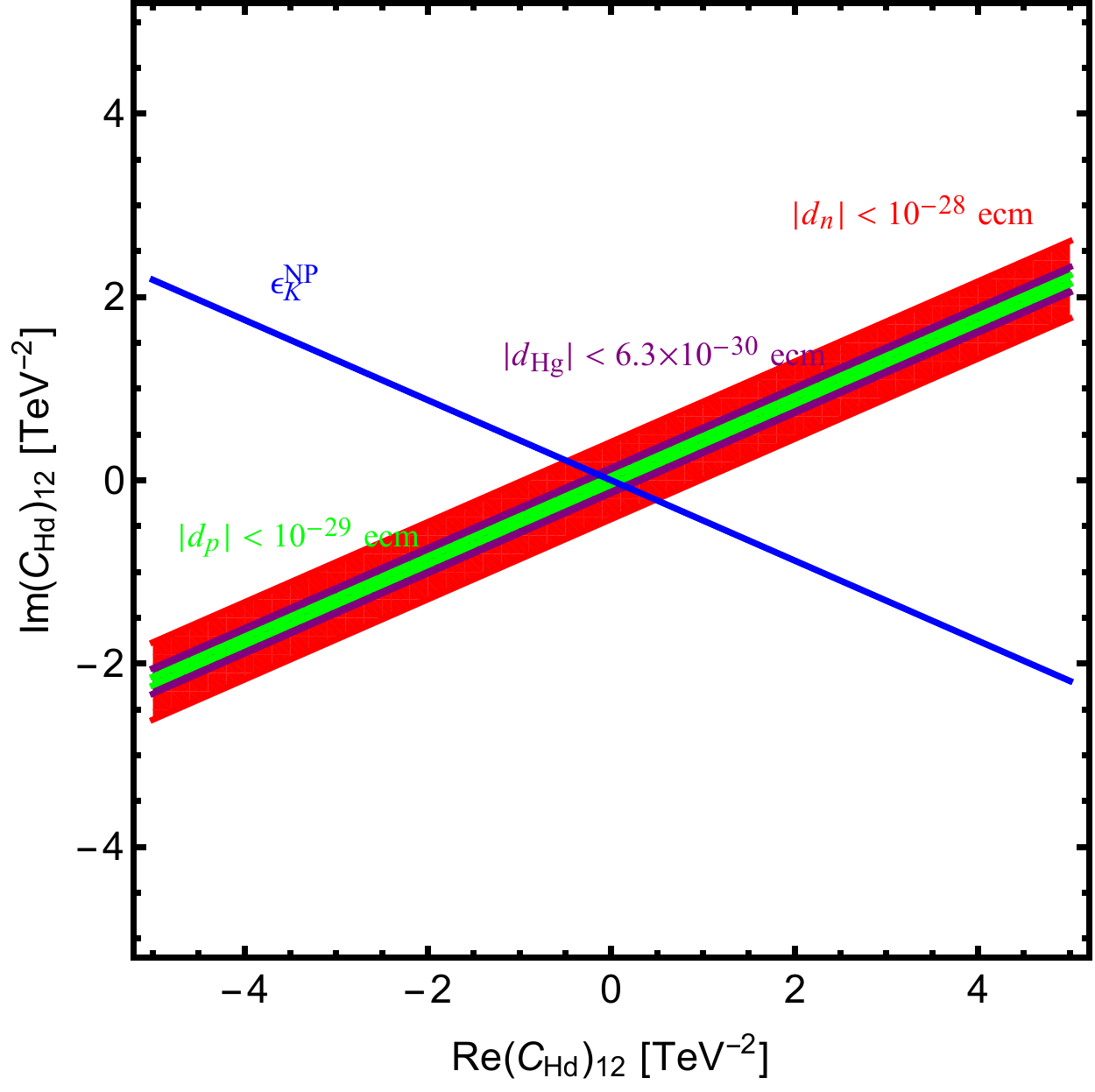}\hspace{25mm}
\includegraphics[scale=0.65, bb= 0 0 256 254]{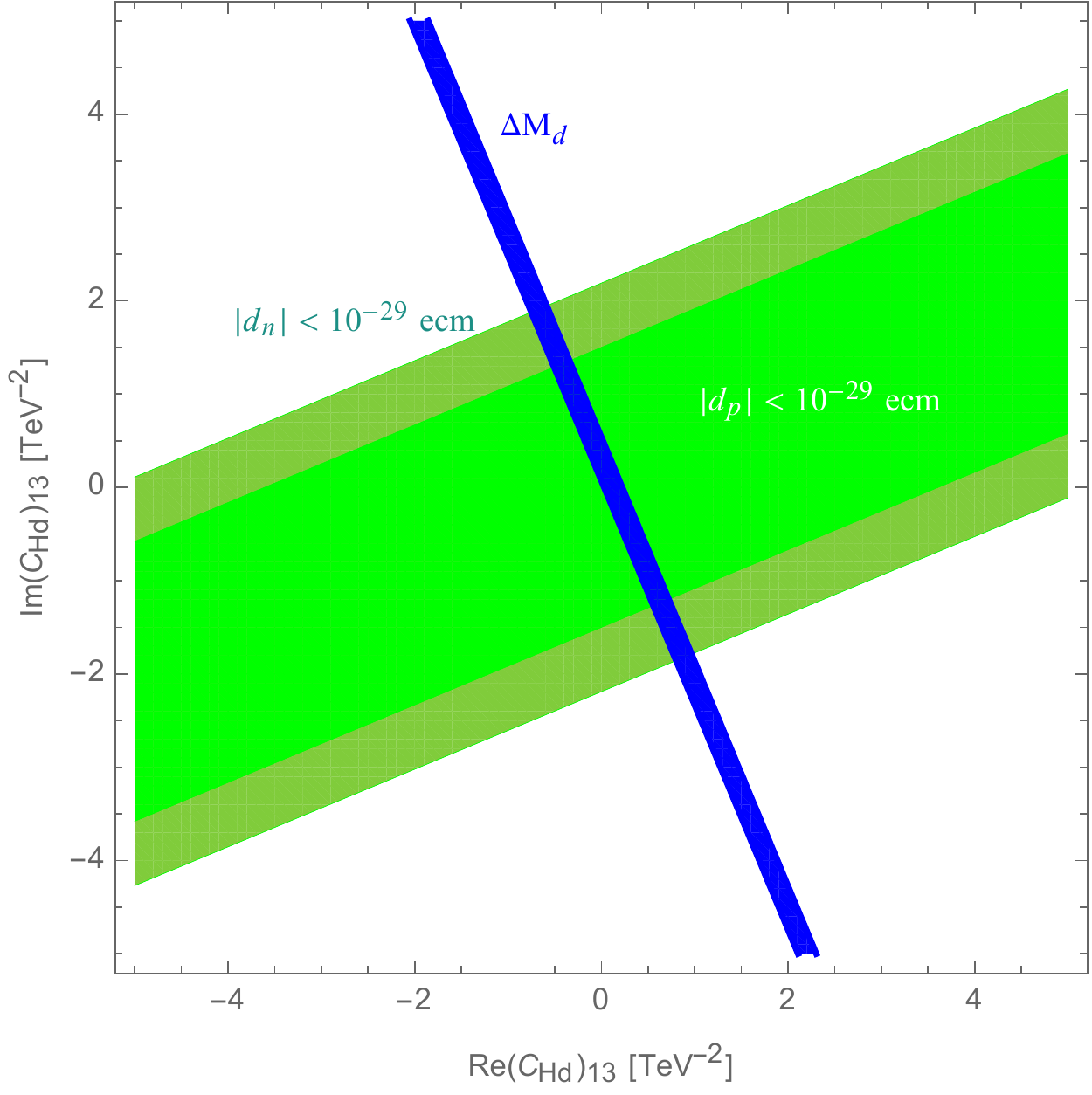}
\end{center}
\caption{
Same as Fig.~\ref{fig:EDM_f_C1qd}, but the deep green region in the right panel is $|d_n| < 10^{-29}~e\,{\rm cm}$, which is one order of magnitude weaker than the future sensitivity.
}
\label{fig:EDM_f_CHd}
\end{figure}
%%%%%%%%%%%%%%%%%%%%%%%%%%%%%%%%%%%%%%%%%%%%
%%%%%%%%%%%%%%%%%%%%%%%%%%%%%%%%%%%%%%%%%%%%
\begin{figure}[h]
\begin{center}
\includegraphics[scale=0.65, bb= 110 0 255 256]{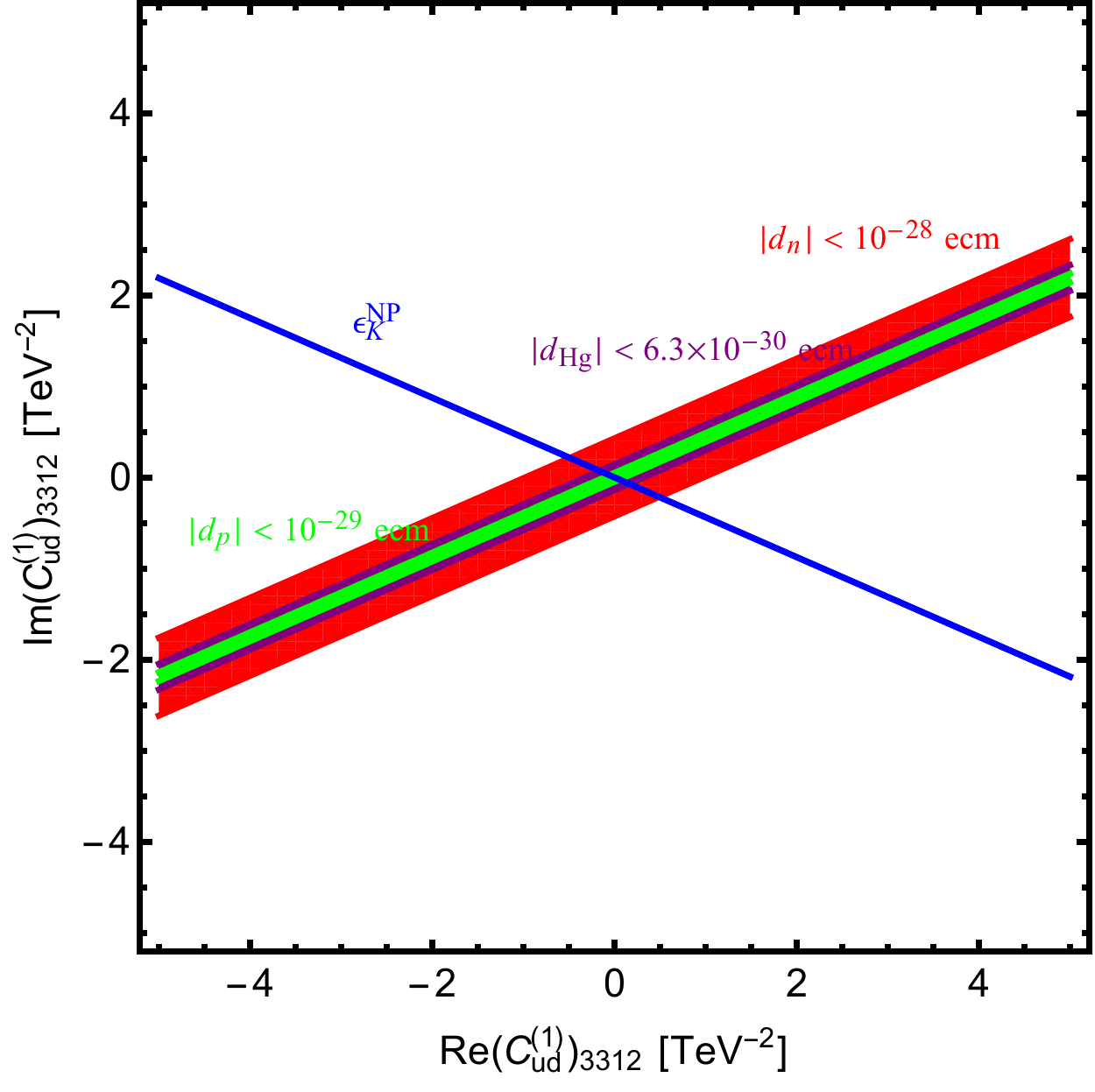}\hspace{25mm}
\includegraphics[scale=0.65, bb= 0 0 255 254]{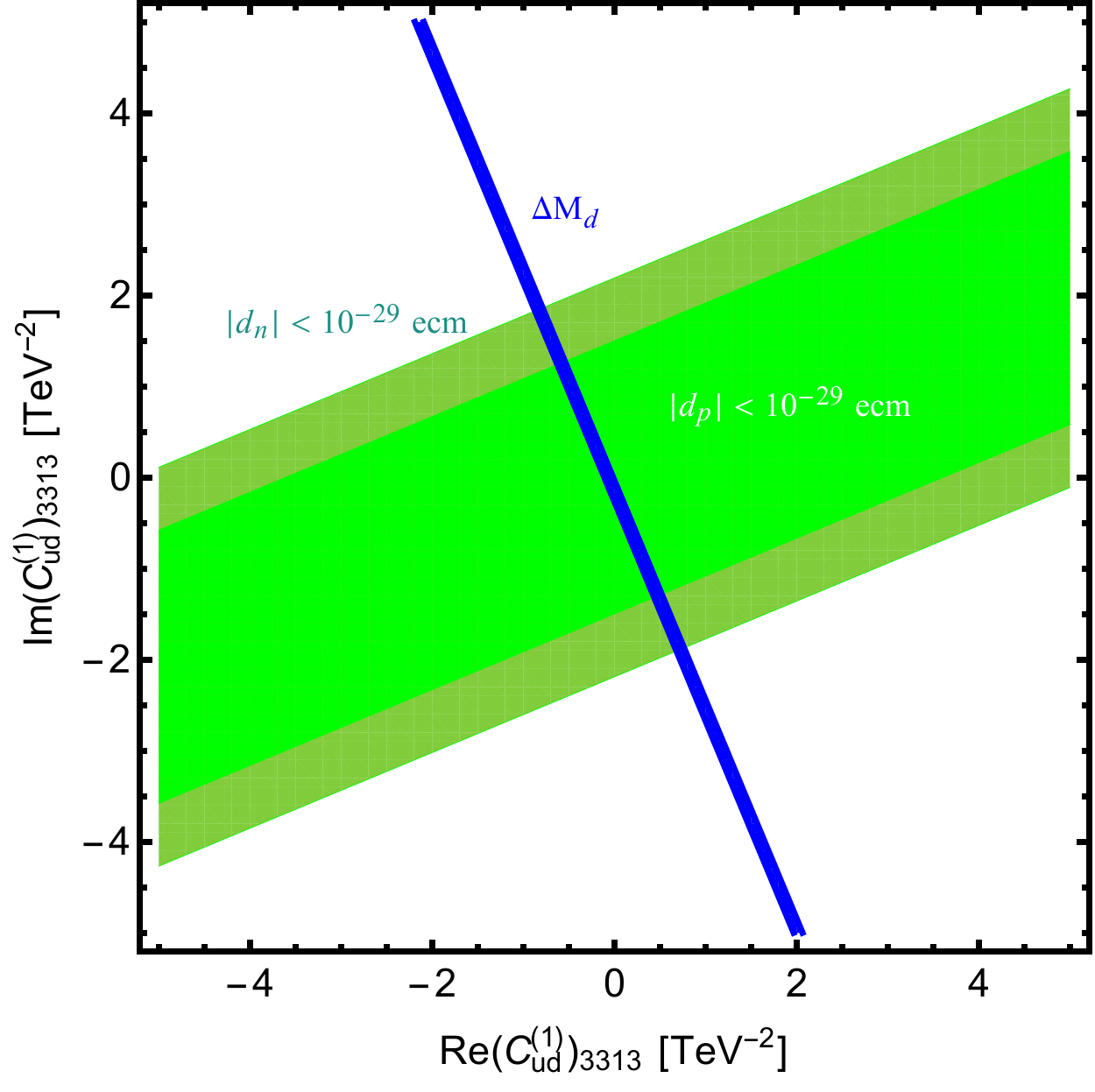}
\end{center}
\caption{
Same as Fig.~\ref{fig:EDM_f_C1qd}, but the deep green region in the right panel is $|d_n| < 10^{-29}~e\,{\rm cm}$, which is one order of magnitude weaker than the future sensitivity.}
\label{fig:EDM_f_C1ud}
\end{figure}
%%%%%%%%%%%%%%%%%%%%%%%%%%%%%%%%%%%%%%%%%%%%

Let us study correlations between the EDMs and the $\Delta F=2$ observables.
The results depend on the SMEFT operators.
The $\Delta S=1$ operators of $(C^{(1,8)}_{qd})_{3312}$, $(C^{(1,8)}_{ud})_{3312}$ and $(C_{Hd})_{12}$ contribute to the EDMs and $\epsilon_K$ via radiative corrections.
Similarly, the $\Delta B=1$ operators of $(C^{(1,8)}_{qd})_{3313}$, $(C^{(1,8)}_{ud})_{3313}$ and $(C_{Hd})_{13}$ affect $\Delta M_{B_d}$ as well as the EDMs.
In Figs.~\ref{fig:EDM_f_C1qd}--\ref{fig:EDM_f_C8ud}, the EDMs and the $\Delta F=2$ observables are estimated for each operator. 
Here, the real and imaginary parts of each Wilson coefficient are varied at the NP scale of 1\,TeV, while the other coefficients are set to be zero at this scale.
In the plots, the current limits from $\epsilon_K$ and $\Delta M_{B_d}$ are drawn by the blue band, where the region inside the band is allowed at the $2\sigma$ level.
On the other hand, contours of the neutron, proton and $^{199}$Hg EDMs are shown by the bands with different colors. 

From the figures, it is noticed that the $^{199}$Hg EDM gives a bound on the $\Delta S=1$ operators, and the proton EDM can provide a better sensitivity for them.
For some of the $\Delta B=1$ operators especially $(C^{(8)}_{qd})_{3313}$ and $(C^{(8)}_{ud})_{3313}$, future measurements of the proton EDMs will also be able to compete with the constraint from $\Delta M_{B_d}$. 
We want to emphasize that the parameter dependence of the EDMs is different from that of $\epsilon_K$.
Thus, the NP contributions to the effective operators can be specified by combining the EDMs with the flavor observables.

Next, let us consider $C^{(1,3)}_{qq}, C^{(1,8)}_{qu}$, and $(C^{(1)}_{Hq})_{12,13}$.
We found that they do not contribute to the EDMs because of the Lorentz structures of these operators.     
In fact, they generate only the vector-type operators of the four quarks below the EWSB scale, which do not violate the $CP$ symmetry. 

Similarly, the operators of $(C^{(3)}_{Hq})_{12,13}$ do not contribute to the EDMs through the four-quark operators.
Let us consider another contribution. 
It is noticed that these operators include $W$ boson interactions by taking the Higgs VEV as
\begin{align}
(H^{\dagger} i\overleftrightarrow{D}_{\mu}^I H ) (\bar{q}^i \gamma^{\mu}\tau^I q^j) &= 
i v^2 \bigg[ 
  (\bar{u}^i \gamma^{\mu} P_L d^j) 
\bigg(  \frac{\sqrt{2}}{v} \partial_{\mu} G^+ -i \frac{g_2}{\sqrt{2}} W^+_{\mu} \bigg)
\notag \\ & \qquad
+ (\bar{d}^i \gamma^{\mu} P_L u^j) 
\bigg( -\frac{\sqrt{2}}{v} \partial_{\mu} G^- -i \frac{g_2}{\sqrt{2}} W^-_{\mu} \bigg)
\bigg] + \ldots
\label{chq3in}
\end{align}
in the Feynman-'t Hooft gauge, where $G^{\pm}$ is the NG bosons.
Here, all the quark fields are left-handed in these interactions.
Then, they seem to generate the electric and chromoelectric dipole moments through penguin diagrams of the $W$ boson loops. 
However, it can be checked that such contributions vanish by paying attention to the chirality structure of the quark. 
Hence, the operators of $(C^{(3)}_{Hq})_{12,13}$ do not contribute to the nucleon EDMs.

Finally, let us comment on $C_{dd}$.
This operator can also contribute to the EDMs through the RGEs and matching conditions.
However, these contributions are found to be very small, and we do not discuss them anymore.

%%%%%%%%%%%%%%%%%%%%%%%%%%%%%%%%%%%%%%%%%%%%%%%%%%%%%%%%
\section{Conclusions}
%%%%%%%%%%%%%%%%%%%%%%%%%%%%%%%%%%%%%%%%%%%%%%%%%%%%%%%%

We studied the nuclear EDMs induced by the SMEFT $\Delta F=1$ operators and their correlations with the $\Delta F=2$ observables. 
These SMEFT operators contribute to them through the $W$ boson loops. 
The radiative corrections via the RGEs and the matching conditions at the EWSB scale are taken into account. 
In particular, we provide the one-loop formulae of the matching conditions for the EDMs.

It was found that some of the operators are already excluded for $M_{\rm NP} \lesssim 1\text{--}9\,{\rm GeV}$ by the $^{199}$Hg EDM, and future experiments for the proton EDM may be able to probe those in $M_{\rm NP} \lesssim 2\text{--}10\,{\rm TeV}$. 
Compared with $\epsilon_K$ and $\Delta M_{B_d}$, it was shown that the nuclear EDMs can provide a complementary information on the $\Delta F=1$ effective operators in future. 

Other nuclear EDMs such as $^{129}$Xe and $^{225}$Ra can also be sensitive to the $CP$-violating baryon-meson interactions.
Although the current bounds are weaker than that of $^{199}$Hg, they would be examined better in future experiments~(see e.g., Ref.~\cite{Strategy:2019vxc}).
Although their theoretical calculations suffer from potentially large uncertainties in estimating the the Schiff moment, it is interesting to study future sensitivities to the SMEFT $\Delta F=1$ operators, which will be explored elsewhere.  

{\it Note added}:
while we are submitting this letter, a new article \cite{Bertolini:2019out} was published on arXiv;
the authors argued that an enhancement factor coming from the strange quark mass which was mentioned in Ref.~\cite{Haba:2018byj} and is quoted in Eq.~\eqref{eq: Aeq} disappears. 
Since this factor can induce a large contribution to the neutron and proton EDMs, the numerical analysis in this article may be affected.

%%%%%%%%%%%%%%%%%%%%%%%%%%%%%%%%%%%%%%%%%%%%%%%%%%%%%%%%
\vspace{1em}
\noindent {\it Acknowledgements}:
This work was supported by JSPS KAKENHI No.~16K17681 (M.E.) and 16H03991 (M.E.).
We are indebted to the referee for important comments on an earlier version of this letter.

\appendix
%%%%%%%%%%%%%%%%%%%%%%%%%%%%%%%%%%%%%%%%%%%%%%%%%%%%%%%%
\section{Estimation of neutron and proton EDMs}
\label{sec:EDMchi}
%%%%%%%%%%%%%%%%%%%%%%%%%%%%%%%%%%%%%%%%%%%%%%%%%%%%%%%%
In this section, let us explain how to estimate the contributions of the four-quark operators to the neutron and proton EDMs with the chiral Lagrangian technique~\cite{deVries:2012ab}.
In particular, we follow the analysis explored in Ref.~\cite{Haba:2018byj}.

We consider the neutron and proton EDMs through meson condensations induced by the $CP$-violating four-quark operators $\tilde{\mathcal{O}}_1^{q'q}$. 
At the parton level, these operators are represented as
\begin{align}
\mathcal{L}_{\rm CPV} &\supset 
\sum_{q'\neq q,q,q'=u,d,s} \tilde{C}_1^{q'q} \tilde{\mathcal{O}}_1^{q' q}
\notag \\ &=
\sum_{i,j,k,l=u,d,s} \bigg[
  i C_{ijkl}^{LRLR} (\bar{q}_i P_R q_j) (\bar{q}_k P_R q_l) 
+ i C_{ijkl}^{RLLR} (\bar{q}_i P_L q_j) (\bar{q}_k P_R q_l)
\bigg] - (L\leftrightarrow R),
\label{partonfour}
\end{align}
where the coefficients are defined as
\begin{align}
C^{LRLR}_{ijkl} = C^{RLLR}_{ijkl} 
= \sum_{q\neq q'} \tilde{C}_1^{q' q} \delta_{i,q'} \delta_{j,q'} \delta_{k, q} \delta_{l,q}.
\end{align}
Under the chiral rotations of $U(3)_L \times U(3)_R$, we impose the following transformations,
\begin{align}
P_L q_i &\to (L)_{ij} P_L q_j,
\\
P_R q_i &\to (R)_{ij} P_R q_j, 
\\
C_{ijkl}^{LRLR} &\to \sum_{m,n,o,p} (L)_{im} (L)_{ko} C_{mnop}^{LRLR} (R^{\dagger})_{nj} (R^{\dagger})_{pl},
\\
C_{ijkl}^{RLLR} &\to \sum_{m,n,o,p} (R)_{im} (L)_{ko} C_{mnop}^{RLLR} (L^{\dagger})_{nj} (R^{\dagger})_{pl},
\end{align}
with $L,R$ are transformation matrices of $U(3)_L$ and $U(3)_R$, respectively.
Then, the right-hand side of Eq.~\eqref{partonfour} is invariant under this transformation. 

By adopting this symmetry in the meson chiral Lagrangian, the $CP$-violating terms are written at $\mathcal{O}(p^2)$ as
\begin{align}
\mathcal{L}_{\rm CPV}^{\rm meson} &= 
\frac{F_{\pi}^2}{4} {\rm Tr} \left[(D_{\mu}U)^{\dagger} D^{\mu} U +\chi (U +U^{\dagger}) \right]
+ \frac{F_0^2 -F_{\pi}^2}{12} 
  {\rm Tr} \left[ U D_{\mu} U^{\dagger} \right] 
  {\rm Tr} \left[ U^{\dagger} D^{\mu} U \right]
\notag \\ &
+ a_0 {\rm Tr} \left[\ln U -\ln U^{\dagger} \right]^2
\notag \\ &
+ \sum_{i,j,k,l=u,d,s} \bigg[ 
i C^{LRLR}_{ijkl} \left(
  c_1 [U]_{ji} [U]_{lk} 
- c_1 [U^{\dagger}]_{ji} [U^{\dagger}]_{lk}
+ c_2 [U]_{li} [U]_{jk} 
- c_2 [U^{\dagger}]_{li} [U^{\dagger}]_{jk} 
\right)
\notag \\ &
+ i C^{RLLR}_{ijkl} \left(
  c_3 [U^{\dagger}]_{ji} [U]_{lk}
- c_3 [U]_{ji} [U^{\dagger}]_{lk} 
\right)
\bigg],
\label{mesonpo}
\end{align} 
where $U,\chi$ are defined as
\begin{align}
U &={\rm exp} \left[\frac{2 i}{\sqrt{6}} \eta_0 I_3 +\frac{2i}{F_{\pi}}\Pi \right],~~~
\chi = 2 B_0 {\rm diag} \left(m_u,m_d,m_s \right),
\\
I_3 &= {\rm diag}\left(1,1,1 \right),~~~
\Pi =
\begin{pmatrix}
\frac{1}{2}\pi^0 +\frac{1}{2\sqrt{3}}\eta_8 & 
\frac{1}{\sqrt{2}}\pi^+ & 
\frac{1}{\sqrt{2}} K^+ \\
\frac{1}{\sqrt{2}}\pi^- & 
-\frac{1}{2}\pi^0 +\frac{1}{2\sqrt{3}}\eta_8 & 
\frac{1}{\sqrt{2}}K^0 \\
\frac{1}{\sqrt{2}}K^- & 
\frac{1}{\sqrt{2}}\bar{K}^0 & 
-\frac{1}{\sqrt{3}}\eta_8
\end{pmatrix}.
\end{align} 
Here, $F_{\pi}$ is the decay constant of the pion, and $F_0$ is that for $\eta_0$.
The mesons matrix $U$ transforms as $U\to R U L^{\dagger}$ under $U(3)_L \times U(3)_R $. 
We approximate as $F_0 \simeq F_{\pi}$, $B_0 \simeq m_{\pi}^2/(m_u +m_d)$ and $48 a_0/F_0^2 \simeq m_{\eta}^2 +m_{\eta'}^2 -2 m_K^2$.
By a naive dimensional analysis, we estimate the unknown low-energy constants, $c_1,c_2$ and $c_3$, as
\begin{align}
c_1 \sim c_2 \sim c_3 \sim \frac{(4\pi F_{\pi})^6}{(4\pi)^4}.
\end{align}

From Eq.~\eqref{mesonpo}, the scalar potential for the neutral mesons, $\pi^0,\eta_8$ and $\eta_0$, is extracted as 
\begin{align}
V(\pi^0,\eta_8,\eta_0) &=
F_{\pi}^2 B_0 \bigg[
m_u \cos \left(
  \frac{\pi^0}{F_{\pi}} 
+ \frac{\eta_8}{\sqrt{3}F_{\pi}} 
+ \frac{2\eta_0}{\sqrt{6}F_0} 
\right)
+ m_d \cos \left(
- \frac{\pi^0}{F_{\pi}} 
+ \frac{\eta_8}{\sqrt{3}F_{\pi}} 
+ \frac{2 \eta_0}{\sqrt{6}F_0} 
 \right)
\notag \\ & \qquad\qquad
+ m_s \cos \left(
- \frac{2 \eta_8}{\sqrt{3} F_{\pi}} 
+ \frac{2 \eta_0}{\sqrt{6}F_0} 
\right)
\bigg]
-24 \frac{a_0}{F_0^2} (\eta_0)^2
\notag \\ &
+ 2 c_1 \bigg[
  \left(\tilde{C}_1^{ud} 
+ \tilde{C}_1^{du} \right) \sin \left(
  \frac{2 \eta_8}{\sqrt{3} F_{\pi}}
+ \frac{4\eta_0}{\sqrt{6}F_0} 
\right)
+ \left( \tilde{C}_1^{us} + \tilde{C}_1^{su} \right) 
\sin \left(
  \frac{\pi^0}{F_{\pi}} 
- \frac{\eta_8}{\sqrt{3}F_{\pi}} 
+ \frac{4 \eta_0}{\sqrt{6}F_0} 
\right)
\notag \\ & \qquad\quad
+ \left( \tilde{C}_1^{ds} + \tilde{C}_1^{sd} \right) 
\sin \left(
- \frac{\pi^0}{F_{\pi}} 
- \frac{\eta_8}{\sqrt{3}F_{\pi}} 
+ \frac{4 \eta_0}{\sqrt{6}} 
\right)
\bigg]
\notag \\ &
+ 2 c_3 \bigg[ 
\left( \tilde{C}_1^{ud} - \tilde{C}_1^{du} \right) 
\sin \left( -\frac{2 \pi^0}{F_{\pi}} \right)
+ \left( \tilde{C}_1^{us} - \tilde{C}_1^{su} \right)
\sin \bigg(
- \frac{\pi^0}{F_{\pi}} 
- \frac{\sqrt{3}\eta_8 }{F_{\pi}} 
\bigg)
\notag \\ & \qquad\quad
+ \left( \tilde{C}_1^{ds} - \tilde{C}_1^{sd} \right) 
\sin \bigg( 
  \frac{\pi^0}{F_{\pi}}
- \frac{\sqrt{3} \eta_8}{F_{\pi}} 
\bigg)
\bigg].
\label{nupo}
\end{align}
Since we are interested only in $\tilde{C}^{ds}_1$ and $\tilde{C}^{sd}_1$, the other Wilson coefficients are set to be zero.
Then, the VEVs of the meson fields are obtained at the leading order of $\tilde{C}^{ds}_1$ and $\tilde{C}^{sd}_1$ as
\begin{align}
\frac{{\langle\pi^0 \rangle}}{F_{\pi}} \simeq& 
- \left( \tilde{C}^{ds}_1 + \tilde{C}^{sd}_1 \right)
\frac{c_1}{B_0 F_{\pi}^2} 
\frac{ B_0 F_{\pi}^2 m_u m_s + 8 a_0  (m_d +2 m_s) }
     { B_0 F_{\pi}^2 m_u m_d m_s + 8 a_0 (m_u m_d +m_d m_s + m_s m_u) } 
\notag \\ &
+ \left( \tilde{C}^{ds}_1 - \tilde{C}^{sd}_1 \right)
\frac{c_3}{B_0 F_{\pi}^2} 
\frac{ B_0 F_{\pi}^2 m_s m_u - 8 a_0 [m_d -2 (m_u +m_s)] }
     { B_0 F_{\pi}^2 m_u m_d m_s + 8 a_0 (m_u m_d +m_d m_s + m_s m_u) },
\\
\frac{{\langle\eta_8 \rangle}}{F_{\pi}} \simeq& 
- \left( \tilde{C}^{ds}_1 + \tilde{C}^{sd}_1 \right)
\frac{c_1}{\sqrt{3} B_0 F_{\pi}^2} 
\frac{ B_0 F_{\pi}^2 m_u (2 m_d -m_s) + 24 a_0 m_d }
     { B_0 F_{\pi}^2 m_u m_d m_s + 8 a_0 (m_u m_d + m_d m_s + m_s m_u) }
\notag \\ &
- \left( \tilde{C}^{ds}_1 - \tilde{C}^{sd}_1 \right)
\frac{c_3}{\sqrt{3} B_0 F_{\pi}^2} 
\frac{ B_0 F_{\pi}^2 (2 m_d +m_s)m_u + 24 a_0 (m_d +2 m_u) }
     { B_0 F_{\pi}^2 m_u m_d m_s + 8 a_0 (m_u m_d + m_d m_s + m_s m_u) },
\\
\frac{{\langle\eta^0 \rangle}}{F_{0}} \simeq&
\left( \tilde{C}^{ds}_1 + \tilde{C}^{sd}_1 \right)
\frac{\sqrt{2}c_1}{\sqrt{3}B_0 F_{\pi}^2}
\frac{ B_0 F_{\pi}^2(m_d +m_s)m_u }
     { B_0 F_{\pi}^2 m_u m_d m_s + 8 a_0 (m_u m_d + m_d m_s + m_s m_u) }
\notag \\ & 
+ \left( \tilde{C}^{ds}_1 - \tilde{C}^{sd}_1 \right)
\frac{\sqrt{2}c_3}{\sqrt{3}B_0 F_{\pi}^2} 
\frac{ B_0 F_{\pi}^2 (m_d -m_s)m_u }
     { B_0 F_{\pi}^2 m_u m_d m_s + 8 a_0 (m_u m_d + m_d m_s + m_s m_u) }.
\end{align}

On the other hand, the baryon chiral Lagrangian is obtained at $\mathcal{O}(p^2)$ as
\begin{align}
\mathcal{L}_{\rm baryons} &=
{\rm Tr} \left[
\bar{B} i\gamma^{\mu} \left( \partial_{\mu} B + [\Gamma_{\mu},B ] \right) 
- M_B \bar{B}B 
\right]
\notag \\ &
- \frac{D}{2} {\rm Tr} \left[ \bar{B} \gamma^{\mu} \gamma_5 \{\xi_{\mu},B \} \right] 
- \frac{F}{2} {\rm Tr} \left[ \bar{B} \gamma^{\mu} \gamma_5 [\xi_{\mu},B ] \right] 
- \frac{\lambda}{2} 
{\rm Tr} \left[ \xi_{\mu} \right] 
{\rm Tr} \left[ \bar{B} \gamma^{\mu} \gamma_5 B \right]
\notag \\ &
+ b_D {\rm Tr} \left[ \bar{B} \{\chi_+ ,B \} \right] 
+ b_F {\rm Tr} \left[ \bar{B} [\chi_+ ,B] \right]
+ b_0 {\rm Tr} \left[ \chi_+ \right] 
      {\rm Tr} \left[ \bar{B} B \right] 
+ \cdots,
\label{eq: baryonin}
\end{align}
where the baryon matrix $B$ is defined as
\begin{align}
B = 
\begin{pmatrix}
  \frac{1}{\sqrt{2}} \Sigma^0 + \frac{1}{\sqrt{6}} \Lambda^0 & 
\Sigma^+ & 
p \\
\Sigma^- & 
- \frac{1}{\sqrt{2}} \Sigma^0 + \frac{1}{\sqrt{6}} \Lambda^0 & 
n \\
\Xi^- & 
\Xi^0 & 
-\frac{2}{\sqrt{6}} \Lambda^0
\end{pmatrix},
\end{align}
with $\xi_{L,R}$ defined as $U=\xi_R \xi^{\dagger}_L$, and $\xi_R =\xi_L^{\dagger}$.
Also, $M_B$ is the baryon mass.
The definitions of $\Gamma_{\mu}$, $\xi_{\mu}$, and $\chi_+$ are
\begin{align}
\Gamma_{\mu} &= 
  \frac{1}{2} \xi^{\dagger}_R \left( \partial_{\mu} - i r_{\mu} \right) \xi_R 
+ \frac{1}{2} \xi^{\dagger}_L \left( \partial_{\mu} - i l_{\mu} \right) \xi_L,
\\
\xi_{\mu} &= 
  i \xi^{\dagger}_R \left( \partial_{\mu} -   r_{\mu} \right) \xi_R 
- i \xi^{\dagger}_L \left( \partial_{\mu} - i l_{\mu} \right) \xi_L,
\\
\chi_+ &= 
  2 B_0 \xi^{\dagger}_L {\rm diag} \left( m_u, m_d, m_s \right) \xi_R 
+ 2 B_0 \xi^{\dagger}_R {\rm diag} \left( m_u, m_d ,m_s \right) \xi_L.
\end{align} 

By inserting the meson VEVs into the baryon chiral Lagrangian, the $CP$-violating baryon-meson interactions become
\begin{align}
\mathcal{L}_{\rm baryons} &\supset 
  \bar{g}_{np\pi^-} \bar{n}p \pi^- 
+ \bar{g}_{ n\Sigma  K^+} \overline{\Sigma^+}p K^+ 
\notag \\ &
+ \bar{g}_{\pi^+ np} \bar{p}n \pi^+ 
+ \bar{g}_{K^+ \Lambda p} \bar{p} \Lambda K^+ 
+ \bar{g}_{K^+ \Sigma^0 p} \bar{p}\Sigma^0 K^+,
\end{align}
where the coupling constants are obtained as
\begin{align}
\bar{g}_{n \Sigma K^+} &=
\frac{B_0}{F_{\pi}} (b_D -b_F) \bigg[
- \frac{1}{\sqrt{2}} (3 m_u +m_s) 
\frac{ {\langle \pi^0\rangle} }{F_{\pi}}
+ \frac{1}{\sqrt{6}} (-m_u + 5 m_s) 
\frac{ {\langle \eta_8\rangle} }{F_{\pi}}
- \frac{4}{\sqrt{3}} (m_u +m_s) 
\frac{ {\langle \eta_0\rangle} }{F_0}
\bigg],\label{eq: Aeq}
\\ 
\bar{g}_{\pi^+ np} &=
\frac{B_0}{F_{\pi}} (b_D +b_F) \bigg[
  \sqrt{2} (m_d-m_u)
\frac{{\langle \pi^0\rangle}}{F_{\pi}}
- \frac{2\sqrt{2}}{\sqrt{3}} (m_u +m_d)
\frac{ {\langle \eta_8\rangle} }{F_{\pi}}
- \frac{4}{\sqrt{3}} (m_u +m_d)
\frac{ {\langle \eta^0\rangle} }{F_0}
\bigg],
\\
\bar{g}_{K^+ \Lambda p} &=
\frac{B_0}{F_{\pi}} (b_D + 3b_F) \bigg[
  \frac{1}{2\sqrt{3}} (m_s + 3m_u) 
\frac{ {\langle \pi^0\rangle} }{F_{\pi}}
+ \frac{1}{6} (m_u -5 m_s) 
\frac{ {\langle \eta_8\rangle} }{F_{\pi}}
+ \frac{2\sqrt{2}}{3} (m_u +m_s) 
\frac{ {\langle \eta^0\rangle} }{F_0}
\bigg],
\\
\bar{g}_{K^+ \Sigma^0 p} &=
\frac{B_0}{F_{\pi}} (b_D-b_F) \bigg[
- \frac{1}{2} (m_s + 3 m_u) 
\frac{ {\langle \pi^0\rangle} }{F_{\pi}}
- \frac{1}{2\sqrt{3}} (m_u -5 m_s) 
\frac{ {\langle \eta_8\rangle} }{F_{\pi}} 
- 2\sqrt{\frac{2}{3}} (m_u +m_s) 
\frac{ {\langle \eta^0\rangle} }{F_0}
\bigg].
\end{align}
These couplings contribute to the neutron and proton EDMs through the baryon-meson loop diagrams.
By following the analysis in Ref.~\cite{Guo:2012vf}, the EDMs are estimated as
\begin{align}
d_n &\sim 
- \frac{e}{8\pi^2 F_{\pi}} \bigg[
  \frac{\bar{g}_{np\pi^-}}{\sqrt{2}} (D+F) 
\left( 1 + \ln \frac{m_{\pi}^2}{m_N^2} \right)
- \frac{\bar{g}_{n \Sigma K^+}}{\sqrt{2}} (D-F) 
\left( 1 + \ln \frac{m_{K^+}^2}{m_N^2} + \frac{\pi (m_{\Sigma^-}-m_n)}{m_{K^+}} \right) 
\bigg],
\label{eq:dn1}
\\
d_p &\sim 
- \frac{e}{8\pi^2 F_{\pi}} \bigg[
- \frac{\bar{g}_{np\pi^-}}{\sqrt{2}} (D+F) \left( 1 + \ln \frac{m_{\pi}^2}{m_N^2} \right)
+ \frac{\bar{g}_{K^+ \Lambda p}}{2\sqrt{3}}(D+3F)
\left( 1 + \ln \frac{m_{K^+}^2}{m_N^2} + \frac{\pi (m_{\Lambda}-m_n)}{m_{K^+}} \right)
\notag \\ &\qquad\qquad~~~~
- \frac{\bar{g}_{K^+ \Sigma^0 p}}{2} (D-F) 
\left( 1 + \ln \frac{m_{K^+}^2}{m_N^2} + \frac{\pi (m_{\Sigma^0}-m_n)}{m_{K^+}} \right)
\bigg],
\label{eq:dp1}
\end{align}
where the finite terms of the leading contributions are shown\footnote{
The loop diagrams are divergent, and the dimensional regularization is adopted.
The renormalizations are discussed in Ref.~\cite{Guo:2012vf}.
}, and the renormalization scale is set to be the nucleon mass, $m_N$.   

Applying the pion decay constant $F_{\pi}=86.8\,{\rm MeV}$~\cite{Borsanyi:2012zv}, the meson-baryon couplings $D=0.804$ and $F=0.463$ from the hyperon $\beta$ decays~\cite{Abele:2003pz}, the low-energy constants $b_D=0.161\,{\rm GeV}^{-1}$ and $b_F =-0.502\,{\rm GeV}^{-1}$ from the baryon octet mass splittings~\cite{Alarcon:2012nr}, and the quark masses $m_u (1\,{\rm GeV}) =2.791\,{\rm MeV}, m_d (1\,{\rm GeV})=5.754\,{\rm MeV}$ and $m_s (1\,{\rm GeV})=116.9\,{\rm MeV}$ evaluated with the QCD four-loop RGEs~\cite{Chetyrkin:2000yt} from the lattice result $m_{ud}(2\,{\rm GeV})=3.364\,{\rm MeV}$, $m_u/m_d=0.485$ and $m_s(2\,{\rm GeV})=92.03\,{\rm MeV}$~\cite{Aoki:2019cca}, we finally obtain the neutron and proton EDMs as
\begin{align}
\frac{d_n}{e} &\sim \left( -0.026 \tilde{C}_1^{ds} + 0.169 \tilde{C}_1^{sd} \right){\rm GeV^{-1}},
\\
\frac{d_p}{e} &\sim \left(  0.023 \tilde{C}_1^{ds} - 0.149 \tilde{C}_1^{sd} \right){\rm GeV^{-1}}.
\end{align}

%%%%%%%%%%%%%%%%%%%%%%%%%%%%%%%%%%%%%%%%%%%%%%%%%%%%%%%% 

\end{document}